\documentclass{aa}
\usepackage{psfig}
\usepackage{graphicx}
\usepackage{latexsym}
\usepackage{amssymb}
\usepackage{txfonts}
\def\etal{et al.}

\def\teff{\ifmmode T_{\rm eff} \else $T_{\mathrm{eff}}$\fi}

\def\ltsima{$\buildrel<\over\sim$}
\def\lsim{\lower.5ex\hbox{\ltsima}}
\newcommand{\hi}{H~{\sc i}}

\newcommand{\ha}{\ifmmode {\rm H}\alpha \else H$\alpha$\fi}
\newcommand{\hb}{\ifmmode {\rm H}\beta \else H$\beta$\fi}
\newcommand{\lya}{\ifmmode {\rm Ly}\alpha \else Ly$\alpha$\fi}

\def\kms{km s$^{-1}$}

\def\msun{\ifmmode M_{\odot} \else M$_{\odot}$\fi}
\def\msunyr{\ifmmode M_{\odot} {\rm yr}^{-1} \else M$_{\odot}$ yr$^{-1}$\fi}
\def\zsun{\ifmmode Z_{\odot} \else Z$_{\odot}$\fi}
\def\lsun{\ifmmode L_{\odot} \else L$_{\odot}$\fi}

\def\mup{\ifmmode M_{\rm up} \else M$_{\rm up}$\fi}
\def\mlow{\ifmmode M_{\rm low} \else M$_{\rm low}$\fi}

\def\aap{A\&A}

\def\aas{A\&AS}
\def\aj{AJ}
\def\apj{ApJ}
\def\apjl{ApJL}
\def\apjs{ApJS}
\def\mnras{MNRAS}

\newcommand{\oh}{\ifmmode 12 + \log({\rm O/H}) \else$12 + \log({\rm
O/H})$\fi}
\def\flyf{\ifmmode f_{\rm Lyf} \else $f_{\rm Lyf}$\fi}
\def\pz{\ifmmode P(z) \else $P(z)$\fi}
\def\ki2{\ifmmode \chi^2 \else $\chi^2$\fi}
\def\zphot{\ifmmode z_{\rm phot} \else $z_{\rm phot}$\fi}

\newcommand{\xphot}{\ifmmode x_\gamma \else $v_\gamma$\fi}
\newcommand{\xobs}{\ifmmode x_{\rm obs} \else $x_{\rm obs}$\fi}
\newcommand{\xcmf}{\ifmmode x_{\rm CMF} \else $x_{\rm CMF}$\fi}
\newcommand{\vexp}{\ifmmode V_{\rm exp} \else $V_{\rm exp}$\fi}
\newcommand{\vmax}{\ifmmode V_{\rm max} \else $V_{\rm max}$\fi}
\newcommand{\nh}{\ifmmode N_{\rm H} \else $N_{\rm H}$\fi}

\begin{document}
\title{3D \lya\ radiation transfer. I. Understanding \lya\ line
  profile morphologies} 

\author{Anne Verhamme\inst{1}, Daniel Schaerer\inst{1,2}, 
Antonella Maselli\inst{3}
}
\offprints{daniel.schaerer@obs.unige.ch}
\authorrunning{Verhamme, Schaerer \& Maselli}
\titlerunning{3D \lya\ radiation transfer}

\institute{
Observatoire de Gen\`eve,
51, Ch. des Maillettes, CH-1290 Sauverny, Switzerland
\and
Laboratoire d'Astrophysique (UMR 5572),
Observatoire Midi-Pyr\'en\'ees,
14 Avenue E. Belin, F-31400 Toulouse, France
\and
Max-Planck-Institut f\"ur Astrophysik, 
Karl-Schwarzschild-Strasse 1, 
D-85748 Garching, Germany
}
\date{Received date; accepted date}

\abstract{}{
The development of a general code for 3D \lya\ radiation transfer in
galaxies to understand the diversity of \lya\ line profiles observed
in star forming galaxies and related objects.}  
{
Using a Monte Carlo technique we have developed a 3D \lya\ radiation
transfer code allowing for prescribed
arbitrary hydrogen density, ionisation, temperature structures, and dust
distributions, and arbitrary velocity fields and UV photon sources.}
{
As a first test and application we have examined  the \lya\ line
profiles predicted for several simple geometrical configurations and
their dependence on the main input parameters. 
Overall, we find line profiles reaching from doubly peaked symmetric
emission to symmetric Voigt (absorption) in static configurations with
increasing dust content, and asymmetric red- (blue-) shifted emission
lines with a blue (red) counterpart ranging from absorption to
emission (with increasing line/continuum strength) in expanding
(infalling) media. 
In particular we find the following results which are of interest
for the interpretation of \lya\ profiles from galaxies.
{{\em 1)}} Standard \lya\ absorption line fitting of global spectra of
galaxies may lead to an underestimate of the true hydrogen column
density in certain geometrical conditions.
{{\em 2)}} Normal (inverted) P-Cygni like \lya\ profiles
can be obtained in expanding (infalling) media from objects without
any intrinsic \lya\ emission, as a natural consequence of radiation
transfer effects. 
{{\em 3)}} The formation and the detailed shape of \lya\ profiles resulting
from expanding shells has been thoroughly revised.
In particular we find that, for sufficiently large column densities 
(\nh\ $\ga 10^{20}$ cm$^{-2}$), the position of the main \lya\
emission peak is quite generally redshifted by approximately twice the
expansion velocity. 
This is in excellent agreement with the observations of $z \sim 3$ LBGs,
which show that \lya\ is redshifted by  $\sim 2 \vexp$, where $\vexp$
is the expansion velocity measured from the interstellar absorption
lines blueshifted with respect to the stellar redshift. 
This finding indicates also that large scale, fairly symmetric shell
structures must be a good description for the outflows in LBGs.
}{}


\maketitle
\section{Introduction}
\label{s_intro}

The \lya\ line plays an important role in a variety of astrophysical
problems, especially as a diagnostic tool to observe and study the high
redshift universe. 
It is a simple redshift indicator for distant galaxies,  
a frequently used star formation rate diagnostic at high $z$,
as well as an important tool probing the ionisation state of the
intergalactic medium and hence the reionisation epoch.
The \lya\ emission line is e.g.\ a strong feature observed in nearby
star forming galaxies, distant Lyman break galaxies (LBGs), sub-mm
galaxies,  emission line selected galaxies (LAE, for Lyman-$\alpha$
emitters), and in the enigmatic so called Lyman-$\alpha$ blobs (LABs)
whose nature remains debated (e.g.\ Steidel \etal\ 2000, Dijkstra
\etal\, 2005b). 

Since the early suggestion of strong \lya\ emission from young high
redshift galaxies by Partridge \& Peebles (1967)
and until the late 1990s, only few \lya\ emitters have been found 
(cf.\ Djorgovski \& Thompson 1992). 
This lack of \lya\ emission detection has triggered a variety of
studies discussing the possible physical effects (mostly metallicity, dust, 
neutral gas kinematics, and geometry) which may significantly affect and
suppress the \lya\ emission and the resonance line radiation transfer, 
thereby reducing the observed \lya\ intensity and destroying simple expected
correlations, e.g.\ between \lya\ intensity and metallicity,  \lya\ intensity 
and UV continuum flux and others
(Meier \& Terlevich 1981, Hartmann \etal\ 1988, Neufeld 1990, 
Charlot \& Fall 1993, Valls-Gabaud 1993, Kunth \etal\ 1998, Tenorio-Tagle
\etal\ 1999,  Mas-Hesse \etal\ 2003).

In the last few years, with the availability of deeper and wider
surveys such as the Large Area Lyman Alpha (LALA) survey and the
Subaru Deep Field survey, many emission galaxies have been detected
(cf.\ Hu \etal\ 1998, 2004; Kudritzki \etal\ 2000, Rhoads \etal\ 2000,
Ouchi \etal\ 2003, Taniguchi \etal\ 2005).  
Although the majority of these distant \lya\ emitters shows rather
simple asymmetric line profiles, the overall diversity of the observed
\lya\ line shapes, both from star forming galaxies in the nearby
universe and at high-$z$, is quite heterogeneous and complex.
The observed line profiles include schematically pure Voigt absorption
profiles, P-Cygni profiles, double peak profiles, pure (symmetric)
emission line profiles, and combinations thereof (see e.g.\ Kunth
\etal\ 1998, Mas-Hesse \etal\ 2003, Shapley \etal\ 2003, M\"oller
  \etal\ 2004, Venemans \etal\ 2005, Wilman \etal\ 2005,  Noll \etal\
2004, Tapken \etal\ 2004, Tapken 2005). 

Although in principle the main physical processes shaping the \lya\ line 
are known, in practice the inferences drawn so far from \lya\ observations
rely mostly on rather simple measurements (e.g.\ line flux) 
or on oversimplified  Voigt-profile fits, which often have no strong
physical motivations. 
For example, for a \lya\ line profile formed purely within a galaxy
(i.e.\ neglecting subsequent alterations from the intergalactic medium
and/or intervening clouds), it is physically inconsistent to fit one
or several Voigt profiles without making strong implicit assumptions
on the geometry of the neutral gas. 
In such a case, due to the resonance scattering nature of \lya\,
it is in fact unlikely that the resulting emergent line profile 
is actually a Voigt profile and a detailed radiation transfer
calculation has thus to be carried out to predict the proper shape of the
emergent resonance line profile.
In general, quantitative simulations for appropriate geometries and
gas kinematics, which take properly into account the main physical
processes of \lya\ line formation and radiation transfer are therefore
needed for a better understanding of the variety of observed \lya\
line profile morphologies.   
The physical properties of the \lya\ emission mechanism and the ones of the 
ambient gas hosting the \lya\ emitter, which both shape the observed \lya\
line, must be properly modeled to investigate the correspondent impact
on the emergent line profile and their possible degeneracies. This in
turn will provide useful hints to guide the interpretation of the
observed profiles and to gain insight on the physical properties of
the \lya\ emitters and of their environment.  

Analytic solutions for the \lya\ radiation transfer problem have been derived
for simple geometries. 
Neufeld (1990) has extensively studied the case of static plane parallel
slabs, yielding important insight on the line formation mechanism 
and providing solutions for configuration including dust, Bowen fluorescence
and \lya\ pumping of H$_2$ Lyman band lines.
The case of a static, uniform sphere has been recently studied by
Dijkstra \etal\ (2005a). 
Loeb \& Rybicky (1999) and Rybicky \& Loeb (1999) have derived solutions
for \lya\ scattering in a Hubble flow. However, more general geometries and
velocity fields do not allow for an analytic solution and require
alternative approaches. 

Over the last few years, several groups have developed new numerical
algorithms, mostly based on Monte Carlo techniques (Spaans 1996, Ahn
\etal\ 2001, 2002; 2003, Ahn 2004; Zheng \& Miralda-Escud\'e 2002;
Richling \etal\ 2001; Richling 2003; Kobayashi \& Kamaya 2004;
Cantalupo \etal\ 2005; Dijkstra \etal\ 2005ab; Hansen \& Oh 2006;
Tasitsiomi 2006).  
Some of these codes have been specifically designed for and can be reliably
applied only to particular configurations: relatively low column
densities (Richling \etal\ 2001; Richling 2003), Hubble flows
(Kobayashi \& Kamaya 2004), 1D geometry (Ahn \etal), spherically
symmetric configurations Dijkstra \etal\ 2005ab); others are strongly
tailored towards cosmological simulations (Cantalupo \etal\ 2005,
Tasitsiomi 2006) and can deal with clumpy/inhomogeneous media (Spaans,
1996; Richling, 2003; Hansen \& Oh 2006). 
In addition, the effect of dust absorption, which is one of the most important
factor affecting \lya\ transmission, is treated only in some of the codes
above.

In any case, none of these studies has attempted to explain systematically
the observed variety of \lya\ line profile morphologies. Furthermore,
none of the above schemes has so far attempted a detailed modeling of
individual galaxies, taking into account the available observational
constraints,  i.e.\ constraints on the stellar populations, the ionised and
neutral interstellar medium, on dust extinction and including their
spatial distribution and kinematics. 
All this lack in the actual theoretical 
modeling needs to be filled up, in order to extract some information from 
the huge reservoir contained in the available observational data.
With these objectives in mind, and with the main aim of 
improving our understanding of \lya\ in both nearby and distant
starburst galaxies, we have developed a general-purpose 3D \lya\ radiation 
transfer code applicable to arbitrary geometries and velocity fields.

In the present paper we provide a description of the code and test its
validity against known solutions and results from other codes reported
in the literature.  
Exploring different geometries, dust-free and dusty media, and different
input spectra (e.g.\ line emission, or continuum + line),
we examine the resulting line profiles and their dependence on various 
physical parameters.
Our immediate goals are to obtain an overview over the possible 
\lya\ line profile morphologies, and to gain physical insight into
the processes governing them. 
Applications to observed galaxies and other simulations will be
presented later.

The remainder of the paper is structured as follows.
A description of the radiation transfer code is given in Sect.\ \ref{s_code}.
Tests of the code and results for simple geometrical configurations
(slabs, infalling/expanding halos) are presented in Sect.\ \ref{s_test}.
In Sect.\ \ref{s_slab} we comment on the formation of damped (Voigt) \lya\ 
profiles and related profiles.
Spherically expanding dust-free or dusty shells are re-examined in 
Sect.\ \ref{s_shell}. 
An overview of the predicted \lya\ line profile morphologies and 
qualitative comparisons with observations is given in Sect.\ \ref{s_discuss}.
Our main conclusion are summarised in  Sect.\ \ref{s_conclude}.

\section{Radiation transfer code}
\label{s_code}

A general 3D radiation transfer code {\em MCLya} allowing for arbitrary
hydrogen density, ionisation \& temperature structures, dust distributions, 
and velocity fields was developed using a Monte Carlo technique.
The input files and the structure of the code have been designed
for a future joint use with the 3D radiation transfer and photoionisation
code CRASH of Maselli \etal\ (2003).
We now summarise the main ingredients and assumptions made
in this code.

\subsection{Geometry}
The present version of the code assumes a 3D cartesian grid
of $n^3$ cells. Typically we adopt $n = 128$.
The relevant quantities describing a 3D structure are the neutral hydrogen 
density distribution, the dust density distribution, 
the temperature distribution, and the velocity field.
These are prescribed by input files.

\subsection{Photon sources}
\lya\ and/or continuum photons are emitted from one or several point
sources. Each source is described by : 
\begin{itemize}
\item its location, 
\item the total number of emitted photons 
\item optionally their emission direction, if not isotropic,
\item the source spectrum (typically monochromatic, a constant photon density
per frequency or wavelength interval, a Gaussian, or combinations thereof).
\end{itemize}

\subsection{Physical processes}
To capture the essentials of radiation transfer in the UV including and 
around the \lya\ line we include three main physical processes, dust
absorption and scattering and the \lya\ line transfer, in the present
version of our code. 
Given the principles of Monte Carlo simulations, other processes can easily 
be included in the future, if desirable.

\subsubsection{\lya\ line transfer}
\label{s_line}
In the whole section, we describe the \lya\ radiative transfer
  equations in a \emph{static} medium. To adapt them to moving media,
  we just convert frequencies to local co-moving frequencies and
  convert them back to the external frame by a Lorentz transformation,
neglecting terms of order $O(v^2/c^2)$.

A \lya\ photon corresponds to the transition between the $n=2$ and 
$n=1$ levels of a hydrogen atom. This is the strongest \hi\ transition, with
an Einstein coefficient given by $A_{21}=6.265\times 10^8$ s$^{-1}$. The
scattering cross-section of a \lya\ photon as a function of frequency
in the rest frame of the hydrogen atom is:
\begin{center} 
\begin{equation}
\sigma_{\nu} =
f_{12}\,\frac{\pi\,e^2}{m_e\,c}\,\frac{\Gamma/4\,\pi^2}{(\nu -
  \nu_0)^2 + (\Gamma/4\,\pi)^2}, 
\end{equation}
\end{center}
where $f_{12}=0.4162$ is the
\lya\ oscillator strength, $\nu_0=2.466\times 10^{15}$ Hz is the line
center frequency, and $\Gamma=A_{12}$ is the damping constant which
measures the natural line width. 

The optical depth $\tau_{\nu}(s)$ of a photon with 
frequency $\nu$ traveling a path of length $s$ is determined by
convolving the above cross-section with the velocity distribution
characteristic of the absorbing gas, and is of the form:
\begin{equation}
\tau_{\nu}(s) = \int_{0}^{s} \int_{-\infty}^{\infty} n(V_z)
\, \sigma_{\nu} \, dV_z \, dl,
\end{equation}
where $V_z$ denotes the velocity component along the photon's
direction. Thermal motions of Hydrogen are described by a Maxwellian
distribution of atoms velocities whose velocity dispersion, 
$V_{th} = (2\,k_B\,T/m_{\rm{H}})^{1/2} = 12.85 T_4^{1/2}$ \kms, 
corresponds to the Doppler frequency width $\Delta\nu_D = (V_{th}/c)\nu_0$.
Here $T_4$ is the gas temperature in units of $10^4$ K. 
In certain cases an additional turbulent motion, characterised by 
$V_{\rm turb}$, is taken into account in the Doppler parameter $b$ given by
\begin{equation}
b = \sqrt{V_{th}^2 + V_{turb}^2}.
\label{e_b}
\end{equation}

Let us now introduce some useful variables. First the frequency
shift in Doppler units
\begin{equation}
  x = \frac{\nu - \nu_0}{\Delta \nu_D} = \frac{V}{b},
\label{eq_x}
\end{equation}
where the second equation gives the relation between $x$ and a
macroscopic velocity component $V$ measured along the photon propagation
(i.e.\ parallel to the light path and in the same direction).
Second the Voigt parameter 
$a \equiv \frac{\Gamma/4\,\pi}{\Delta\nu_D}=4.7\times10^{-4}\,T_4^{-1/2}$,
or more generally $a=4.7\times10^{-4}\,(12.85\, {\rm km \, s^{-1}}/b)$
for non-zero turbulent velocity.
Adopting this notation, it can be shown that: 
\begin{equation}
\tau_x(s) = \sigma_H(x)\, n_H\, s=
1.041\times 10^{-13}\,T_4^{\phantom{4}-1/2}\, N_H\, \frac{H(x,a)}{\sqrt\pi}
\end{equation}
where $n_H$ is the neutral hydrogen density, and $N_H$ the
corresponding column density. The Hjerting function $H(x,a)$
describes the Voigt absorption profile,
\begin{equation}
H(x,a) = \frac{a}{\pi} \int_{-\infty}^{\infty}
\frac{e^{-y^2}dy}{(y-x)^2+a^2} \approx \left\{\begin{array}{ll}
e^{-x^2} & \textrm{if $|x|<x_c$} \\
\frac{a}{\sqrt\pi x^2} & \textrm{if $|x|>x_c$} \label{s_wing}
\end{array} \right.
\end{equation}
which is often approximated by a central resonant core 
and power-law ``damping wings'' for frequencies below/above a certain 
boundary frequency $x_c$ between core and wings. 
For $a$ in the range of $10^{-2}$ to $10^{-6}$, $x_c$ varies typically
from 2.5 to 4. 
To evaluate $H$ in our code, we use the fit formulae given by Gray (1992).

To characterise the depth of a static medium we will use
$\tau_0$, the optical depth at line center: 
\begin{eqnarray}
\tau_0(s) & = & 1.041\times
  10^{-13}\,T_4^{-1/2}\, N_H\, \frac{H(0,a)}{\sqrt\pi} \\
          & \approx & 3.31\times 10^{-14} \,T_4^{-1/2}\, N_H 
\end{eqnarray}  
for zero turbulent velocity, or more generally
\begin{equation}
\tau_0(s) = 3.31\times 10^{-14} \, (12.85\, {\rm km \, s^{-1}}/b) \, N_H 
\end{equation}
This monochromatic optical depth $\tau_0$ has been used in most recent
studies (e.g.\ Ahn \etal\ 2001, 2002; Zheng \& Miralda-Escud\'e 2002, Hansen
\& Oh 2006); however, it differs from the total \lya\ optical depth
$\tau_0^{\rm Neuf}$ used in the classical work of Neufeld (1990) by
$\tau_0^{\rm Neuf} = \tau_0 \sqrt{\pi}$.

Once the absorption probability is given and before the \lya\ photon
is re-emitted, its frequency and angular distribution must be determined.
If the atom is not perturbed by collisions during the time a \lya\
  photon is absorbed and re-emitted, the frequencies before and after
  scattering are identical in the atom's rest-frame. On the other
  hand, when the atom undergo a collision, the electron is reshuffled
  on another energy level and the frequencies before and after
  scattering are uncorrelated. Given the typically low densities in
  astrophysical media, we assume coherent scattering in our code.

  Concerning the angular redistribution, our code can model the case
  of isotropic as well as the more realistic dipolar redistribution. In
  the case of isotropy, for simplicity and speed we use the angle
  averaged frequency redistribution function $R_{II-A}(x_{\rm
  in},x_{\rm out})$ from Hummer(1962). 
In practice we use pretabulated values of the cumulative frequency
distribution function of $R_{II-A}$ for different input frequencies
  and temperatures. 
In all the static geometries presented in this paper, we use an
  isotropic angular redistribution. As test calculations confirm (cf.\
  Fig~\ref{slab_spec} and also Zheng \& Miralda-Escud\'e 2002 and
  Hansen \& Oh 2006) this is an excellent approximation since the
  \lya\ photons undergo a very large number of scatterings where any
  angular preference is smeared out. 

To avoid numerous core scatterings in
static cases with a high column density, different acceleration
methods have been developed in other radiation transfer codes (e.g.\ Ahn \&
Lee 2002, Djikstra \& Haiman 2005a). In our case, such an acceleration 
is easily included in the redistribution functions by setting artificially
$R_{II-A}(x_{\rm in},x_{\rm out})=\epsilon$ where $\epsilon$ is close to zero 
for $|x_{\rm in}|<x_c$ and  $|x_{\rm out}|<x_c$, which corresponds to
setting the probability to be re-emitted at $|x_{\rm out}|<x_c$ to zero,
when photons are absorbed in the core (i.e.\ when $|x_{\rm in}|<x_c$).
In practice we have not used the acceleration method except for test
cases, as it turns out that all cases shown here are tractable without it.
 
The dipolar angular redistribution has been implemented without
  the use of redistribution functions, but microscopically, following 
the detailed descriptions of former codes (cf.\ Zheng \&
  Miralda-Escud\'e 2002 and Dijkstra \& Haiman 2005a). The necessity to use
  this more physical redistribution is particularly important for
  expanding shells (see section~\ref{s_shell}).

\subsubsection{Dust scattering and absorption}

During its travel in an astrophysical medium, the \lya\ photon will
diffuse on H atoms, but it can also interact with dust: it can either
be scattered, or absorbed. The dust cross-section $\sigma_d$ is
composed of an absorption cross-section $\sigma_a$, and a scattering
cross-section $\sigma_s$:
\begin{equation}
\sigma_d = \sigma_a + \sigma_s
\end{equation}
where $\sigma_{a,s} = \pi\,d^2\,Q_{a,s}$, with $d$ the
typical dust grain size which will affect \lya\ photons, and $Q_{a,s}$
the absorption/scattering efficiency. 
At UV wavelengths the two processes are equally
likely, $Q_a \approx Q_s \approx 1$, so the dust albedo $A=Q_s/(Q_a+Q_s)$ is around 0.5: 
half of the photons interacting with dust will be lost, and half will be
re-emitted in the \lya\ line.

We assume that the dust density $n_{d}$ is proportional to the neutral H
density in each cell
\begin{equation} 
n_{d} = n_H \times \frac{m_{\rm{H}}}{m_d}\,\frac{M_d}{M_{\rm{H}}},
\label{nd}
\end{equation}
where $m_d$ is the grain mass and $m_{\rm{H}}$ the proton mass.
The relevant quantity, $\tau_{d}$ given just below, is described by one
free parameter, the dust to gas ratio $\frac{M_d}{M_{\rm{H}}}$ assuming 
$d=10^{-6}$ cm and $m_d = 3\times 10^{-17}$ g. 
The total (absorption + scattering) dust optical depth seen by a \lya\ photon is then: 
\begin{equation}
\tau_{d} = \tau_{a} +\tau_{s} = \int_{0}^{s} \sigma_{d}\,n_{d}(s) \, ds.
\end{equation}

The relation between the dust absorption optical depth at \lya\ wavelength 
$\tau_a=(1-A)\tau_d$ and the color excess $E_{B-V}$ is given by
\begin{equation}
	E_{B-V} = 1.086 \frac{A_V}{A_{1216} \, R} \, \tau_{a} \approx (0.06 \ldots 0.11) \, \tau_{a}
\end{equation}
where $A_V$ and $A_{1216}$ is the extinction in the $V$ band and at 1216 \AA\
respectively, $R$ the total-to-selective extinction. The smaller numerical
value corresponds to a Calzetti \etal\ (2000) attenuation law for starbursts, 
the larger to the Galactic extinction law from Seaton (1979).

\subsection{Monte Carlo radiation transfer}
For each photon source we emit photons one by one, and we follow each photon 
until escape from our simulation box or absorption by dust. 
Let us now describe one photon's travel.

\subsubsection{Initial emission}
The emission of a photon is characterised by an emission frequency and
direction.  
The frequency $\nu$ (here in the ``external'', i.e.\ observer's frame)
samples the source spectrum representing usually \lya\ line emission
and/or UV continuum photons. 
For media with constant temperature, $\nu$ or more precisely the
emission frequency shift  from the line center, is conveniently
expressed in Dopper units, i.e.\ 
$x = (\nu - \nu_0)/\Delta \nu_D$ (cf.\ Eq.\ \ref{eq_x}).

We assume that the source emission is isotropic (in the local co-moving
frame, if the considered geometry is not static). Thus the emission
direction, described by the two angles $\theta$ and $\phi$, is
randomly selected from
\begin{eqnarray}
\theta & = & cos^{-1}(2\,\xi_1 - 1) \\ 
\phi & = & 2 \, \pi \, \xi_2
\end{eqnarray}
where $\xi_{1,2}$ are random numbers
\footnote{The random numbers generator used in the code is the {\tt ran}
  function from Numerical Recipies in Fortran 90 (Chapter B7, page
  1142)}, $0 \le \xi_{1,2} < 1$. The photon travels in this direction
until it undergoes an interaction. In moving media, the photon
  frequency in the external frame is evaluated by a Lorentz transformation. 

\subsubsection{Location of interaction}

The location of interaction is determined as follows. The optical
depth, $\tau_{int}$, that the photon will travel is determined by sampling
the interaction probability distribution $P(\tau) = 1 - e^{-\tau}$ by setting
\begin{equation}
\tau_{int} = -ln(1-\xi)
\end{equation}
where $\xi$ is another random number.

We sum the optical depth $\tau$ along the photon path $s$: 
\begin{equation}
\tau(s) = \tau_x(s) + \tau_{d}(s)
\end{equation}
and we determine the length $s$ corresponding to $\tau(s) =
\tau_{int}$. We calculate the coordinates corresponding to a travel of
length $s$ in the direction ($\theta,\phi$) starting from the emission
point. This is the location of interaction. Now, we have to compute if
the \lya\ photon interacts with a dust grain or a hydrogen atom.

\subsubsection{Interaction with H or Dust ?}

The probability to be scattered by a hydrogen atom is given by: 
\begin{equation}
P_{\rm{H}}(x) = \frac{n_H \sigma_H(x)}{n_H \sigma_{\rm{H}}(x) + n_{d} \sigma_d},
\end{equation}
where $\sigma_{\rm{H}}(x) =
f_{12}\,\frac{\pi\,e^2}{m_e\,c\,\Delta\nu_D}\,H(x,a)$ is the hydrogen
cross section for a \lya\ photon of frequency $x$. 
We generate a
random number $0 \le \xi < 1$ and compare it to $P_{\rm{H}}$: if
$\xi\,<\,P_{\rm{H}}$, the photon interacts with H, otherwise it is scattered
or absorbed by dust.

\subsubsection{Scattering on H atoms}  
When the photon is absorbed by an H atom, how will it be reemitted ?
 
We first convert the frequency in the external
(observers) frame, $\nu_{ext}$, to the comoving frequency of the fluid,
$\nu_{int}$, with a Lorentz transformation 
\begin{equation}
\nu_{int} = \nu_{ext}(1-\frac{\vec k_i \cdot \vec V}{c}),
\end{equation}
where $\vec k_i$ is the photon direction and $\vec V$ the
macroscopic/bulk velocity of the H atoms.
As already mentioned above, we assume partially coherent scattering
and either isotropic or dipolar angular redistribution. 
After scattering, the new frequency is again converted back to the
external frame. 

\subsubsection{Dust scattering and absorption}

When the photon interacts with dust, we generate a random number
 $0 \le \xi < 1$ determining whether it is absorbed or scattered.
In practice, if $\xi < A= Q_s/(Q_a+Q_s) $ the photon is scattered by dust
and simply reemitted coherently.
Otherwise the photon is absorbed by dust and is considered lost for the 
present simulation.
For the same reasons already discussed above (Sect.\ \ref{s_line})  we
 currently assume that dust scattering is isotropic, but other angular
 distributions (e.g.\ Henyey-Greenstein functions) can easily be implemented.

\subsubsection{Output}

The precedent scheme is repeated until the photon escapes the
simulation volume, or undergoes a dust absorption. Then, we store all
the information concerning this photon in a matrix
and start with the next photon. 
This procedure is repeated for all photons and all emission sources.
Finally all the desired results, such as spatially integrated line profiles, 
monochromatic or integrated \lya\ images, surface brightness contours
for any given line of sight and spatial resolution etc. are computed
from this output matrix. For reasons of symmetry, and to 
maximise the numerical efficiency,
  all the spectra presented hereafter are integrated spectra over all
  directions, except if mentioned otherwise (cf Fig~\ref{slab_cont_ext}).

\begin{figure}[tb]
\includegraphics[height=6cm, width=8cm]{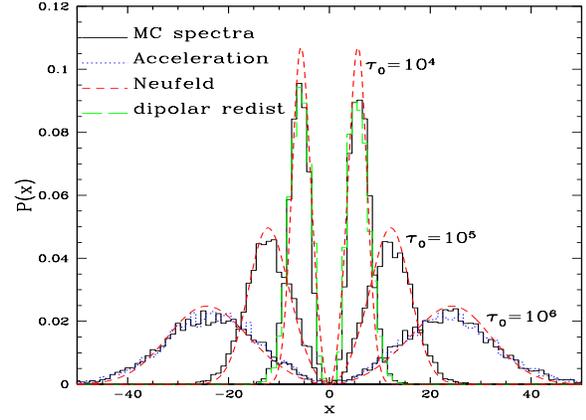}
\caption{Predicted emergent \lya\ profiles for monochromatic line radiation emitted
in a dust-free slab of different optical depths (solid lines) 
compared with analytic solutions from Neufeld (1990, dashed).
The dotted blue curve shows the line profile obtained using a 
frequency redistribution function, which skips a large number of
resonant core scatterings. 
The adopted conditions of the medium are: $T=10$ K (i.e.\ $a=1.5\, 10^{-2}$),
and $\tau_0=10^4$, $10^5$, $10^6$ from top to bottom.
The green long-dashed curve, obtained
  with a dipolar angular redistribution, overlaps perfectly the black
  solid line, obtained with the isotropic angular redistribution
  function, illustrating the fact that 
  in static media, isotropy is a very good approximation.}
\label{slab_spec}
\end{figure}

\begin{figure}[tb]
\includegraphics[height=6cm, width=8cm]{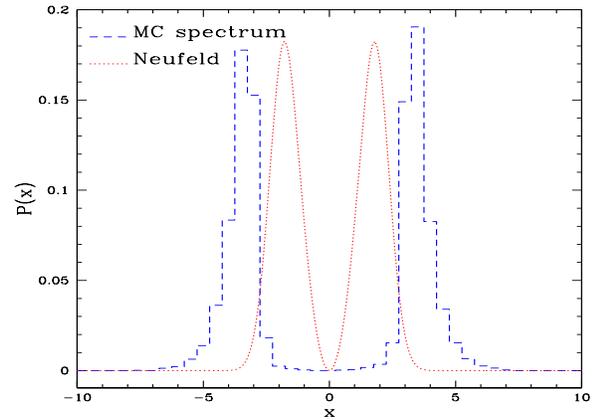}
\caption{Same as Fig.\ \protect\ref{slab_spec} for a slab of small
  optical depth ($\tau_0=10^4$) and  high temperature $T=10^4$ K
  ($a=4.7\, 10^{-4}$) illustrating the deviation from the analytic
  solution of Neufeld (1990) due to the more complete treatment of the
  line absorption profile (see text).}
\label{slab_AhnetLee}
\end{figure}

\section{Validation of the code and examination of simple geometrical
  configurations} 
\label{s_test}

To validate/test our code and to gain insight into basic properties of
the \lya\ radiation transfer we have computed a large number of simulations
for geometrical setups discussed previously in the literature :
  slabs, expanding/infalling halos, expanding shells that we will
  present in details hereafter, and 
  disk-like configurations for which we have perfect agreement with
  Richling \etal\ (2003).

We consider various input spectra (i.e.\ the intrinsic emission line profile
and possible continuum emission), especially the limiting cases of
a pure monochromatic (i.e.\ line) radiation.
Both cases with or without dust are considered.
The case of a source emitting a pure continuum in a plane parallel slab
with and without dust is discussed in Sect.\ \ref{s_slab}.

\subsection{Homogeneous slab}
The best studied case is that of a plane parallel homogeneous slab,
for which analytic solutions of the \lya\ transfer problem
have been derived and discussed in depth by Neufeld (1990).
For a given source position and input spectrum,
the main physical quantities determining the output spectrum
are the temperature of the medium and its optical depth, whatever the
angular redistribution is.
Note that in this section, all profiles have been
  obtained using isotropic angular redistribution, except if the
  contrary is specified.
\subsubsection{Monochromatic radiation, no dust}

As a first test we simulate a dust-free slab with a central plane
source emitting  \lya\ photons in the line center (i.e.\ $\nu_0 =
2.46\times 10^{15}$ Hz or equivalently $x = 0$). 
We choose $T = 10$ K (i.e.\ $a=1.5\times 10^{-2}$), and vary the optical depth
in the line center, $\tau_0 = 10^4,\,10^5,\,10^6$, to be in the
validity range of Neufeld's analytic solution assuming a power-law
absorption profile (Eq.\ \ref{s_wing}): a very optically thick slab,
where $a\,\tau_0\,>(a\,\tau_0)_{\rm crit}$ (cf.\ below).
In this case the emergent \lya\ line profile is given by
(see Neufeld 1990, Eq.\ 2.24):
\begin{equation}
J(\tau_0,x) = \frac{\sqrt6}{24\sqrt\pi} \frac{x^2}{a\tau_0}
\frac{1}{\cosh[\sqrt{\pi^3/54}(|x^3|/a\tau_0)]}.
\end{equation}
As shown in Fig.\ \ref{slab_spec}, our
spectra are in perfect agreement with Neufeld's
predictions\footnote{As already mentioned above our definition of $\tau_0$,
the monochromatic line optical depth at $x=0$ (line center), differs
from Neufeld's definition of $\tau_0^{\rm Neuf}$ which is the total,
i.e.\ frequency integrated, \lya\ optical depth. One has:
$\tau_0^{\rm Neuf} = \tau_0 \sqrt{\pi}$.}.
The spectra are double peaked and symmetric around $x=0$. The 
peak frequency $x_p$ reflects the physical properties of the neutral medium
\begin{equation}
x_p \approx \pm\, 0.88 (a\,\tau_0)^{1/3}.
\label{e_peak}
\end{equation}

As expected, the more optically thick the medium is, the more
separated are the peaks. The width of the peaks becomes larger with
larger $\tau_0$.
In Fig.\ \ref{slab_spec} we also show that computations using our 
acceleration method (cf.\ Sect.\ \ref{s_line}) yield excellent agreement.

For higher temperature, i.e. smaller $a$ (see Fig.\ \ref{slab_AhnetLee})
our spectrum differs from the analytic solution of Neufeld (1990),
as also noticed by Ahn \etal\ (2001) and Zheng \& Miralda-Escud\'e (2002).
This is due to the simplified assumption of a power-law line profile
in the wings, which Neufeld (1990) assumes to be valid for the entire 
absorption line profile $H$ (cf.\ Sect.\ \ref{s_line}). 
As a consequence our peaks are more separated and less symmetric
than expected from Neufeld's calculation. This is due to the fact
that Neufeld's approximation of $H$ underestimates the absorption probability 
in the core, such that photons escape more easily than in the real case and 
their mean escape frequency remains closer to the line center.

For Neufeld's analytical solution, based on the assumption that 
$H(x,a) \approx \frac{a}{\sqrt\pi x^2}$ as in Lorentz wings, to be valid, a minimum
criterium is that the optical depth at
the frequency $x_c$ corresponding to the transition between core and
wings is larger than $1$.
Otherwise, most of the \lya\ photons will escape from the core where Neufeld's
approximation is not valid.
In practice, since $\tau_{\rm crit} = \tau(x_c)=1$ corresponds to 
$(a\tau_0)_{\rm crit}=\sqrt\pi x_c^2$,
the analytic solutions of Neufeld (1990) are valid only for sufficiently
large optical depths, i.e.\ for  $a\,\tau_0\,>(a\,\tau_0)_{\rm crit}$.

\subsubsection{Monochromatic radiation and dust}
\label{s_slab_dust}

We now include dust in the slab and examine its effect on line profiles and
on \lya\ photon destruction.

\begin{figure}[tb]
\includegraphics[height=6cm, width=8cm]{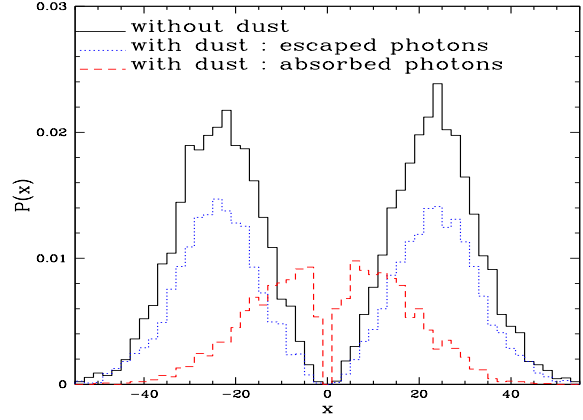}
\caption{Predicted emergent \lya\ profiles for monochromatic line radiation emitted
in a slab of optical depth $\tau_0 =  10^6$, $T=10$ K, and $\tau_a = 10^{-3}$ and 0.
The black solid line shows the emergent profile without dust and 
the blue dotted line the one with dust.
The red dashed line shows the (unobservable) spectrum of the photons
which have been  absorbed by dust.
The effect of dust seems ``grey'' on the emergent line profile because
$\tau_{\rm dust}$ is very weak. For greater $\tau_{\rm dust}$, the inner part
of the peaks is destroyed.}	
\label{slab_dust}
\end{figure}

As illustrated in Fig~\ref{slab_dust}, the presence of dust absorption reduces
the total \lya\ intensity. For the case of monochromatic radiation
with a very small amount of dust shown here, the shape of the emerging \lya\ 
profile remains basically unchanged by dust. 
However, with increasing $\tau_{\rm dust}$, the inner parts of the emission 
peaks are destroyed. 
Although not an observable, we also plot
the spectrum of the absorbed photons (red dotted line on
Fig~\ref{slab_dust}). As expected, it is symmetric with respect to the line
center: it presents two peaks, closer to the line center than those of the emerging
profile. To understand these features, we have to take into
account two competing effects,
the large number of scatterings photons undergo in the line core favouring
a potential interaction with dust, but the low probability to interact with dust
in presence of the very strong HI absorption cross section close to the
line core.
Indeed, the probability to interact with H is around unity in the core
since
\begin{equation}
P_{\rm{H}}(x=0) = \frac{\sigma_{\rm{H}}(0)}{\sigma_{\rm{H}}(0) + 
                 \frac{m_{\rm H}}{n_d} \frac{M_{d}}{M_{\rm H}}\sigma_d}
 \approx 
  \left[ 1 + c^\prime \,T_4^{1/2} \frac{M_d}{M_{\rm H}} \right] ^{-1},
\end{equation}
with $c^\prime = 3.3 \, 10^{-6}$ K$^{-1/2}$
for the dust parameters adopted here.
Therefore none of the photons are absorbed by dust in the very center of the
line. The peaks are located where the probability to interact with
H is lower (for $|x|<x_c$) but where the number of interactions (which ultimately
increase the chances for dust absorption or scattering) is still high, 
i.e.\ on both sides of the line core.

\begin{figure}[tb]
\includegraphics[height=6cm, width=8cm]{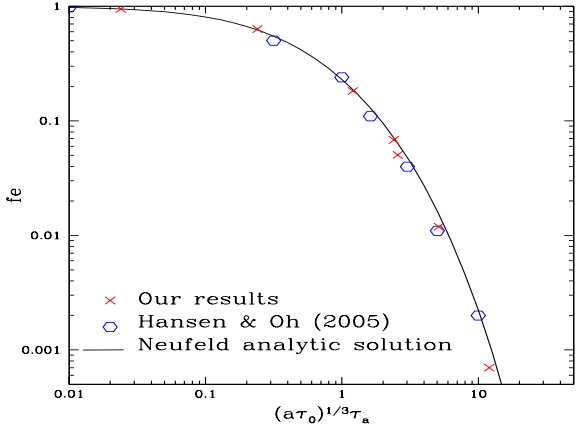}
\caption{Escape fraction $f_e$ for \lya\ line photons emitted
  monochromatically in a dusty slab as a function of $(a\tau_0)^{1/3} \tau_a$.
The excellent agreement between our simulation (crosses), the computations
of Hansen \& Oh (2006) and the approximate analytical solution of
Neufeld (1990) given by Eq.\ \protect\ref{eq_fe} is shown.}
\label{esc_frac_slab}
\end{figure}

For an optically thick slab ($a\tau_0 > 10^3$), Neufeld (1990) gave an
analytic solution to determine the escape fraction, i.e.\ the ratio
between the number of photons which escape the medium and the total
number of emitted photons. 
This escape fraction $f_e$ depends on the combination
$(a\tau_0)^{1/3} \tau_a$, where $\tau_a$ is the optical depth of
absorption from the center to the surface of the slab,
i.e.\ $\tau_a = \tau_d / 2$ here since we consider 
a dust albedo $A \sim 0.5$. 
For a central source and in the limit $(a\tau_0)^{1/3} \gg \tau_a$, 
Neufeld (1990, his Eq.\ 4.43) has derived an approximate expression for $f_e$,
which, in our notation, is
\begin{equation}
f_e = 1/\cosh\left[\frac{\sqrt3}{\pi^{5/12} \zeta} \left[(a\tau_0)^{1/3} \tau_a\right]^{1/2} \right]
\label{eq_fe}
\end{equation}
where $\zeta \approx 0.525$ is a fitting parameter.  
In Fig.~\ref{esc_frac_slab}, we compare our results with this
analytical curve and with the results from Monte-Carlo simulations of 
Hansen \& Oh (2006). Our results are in good agreement with both of them.

\subsubsection{Continuous input spectrum}
The case of a source emitting a pure continuum in a plane parallel slab
with and without dust is discussed in Sect.\ \ref{s_slab}.

\subsection{Expanding/infalling halos}

We now simulate spherical clouds (``halos'') of uniform density, not only static,
but also expanding or collapsing. An isotropic \lya\ source is either located at the
center of the sphere, or uniformly distributed over the whole
volume. In this section, the angular redistribution considered is
  isotropic, after verification  that the emergent profiles are very similar,
  even in the cases with motion.

\subsubsection{Monochromatic source}

\begin{figure*}[hbt]
\begin{tabular}{ll}
\includegraphics[height=6cm, width=8cm]{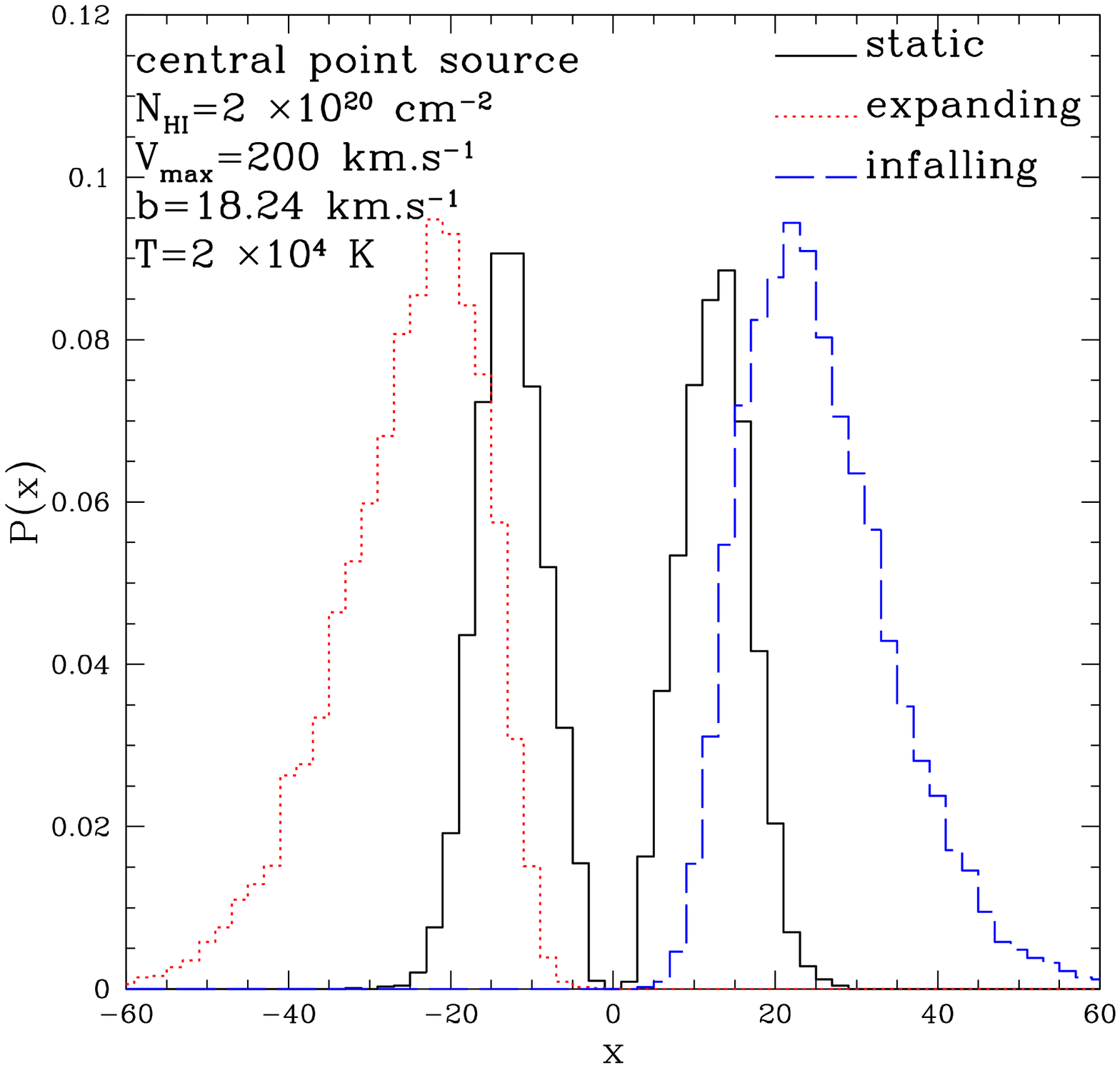} & 
\includegraphics[height=6cm, width=8cm]{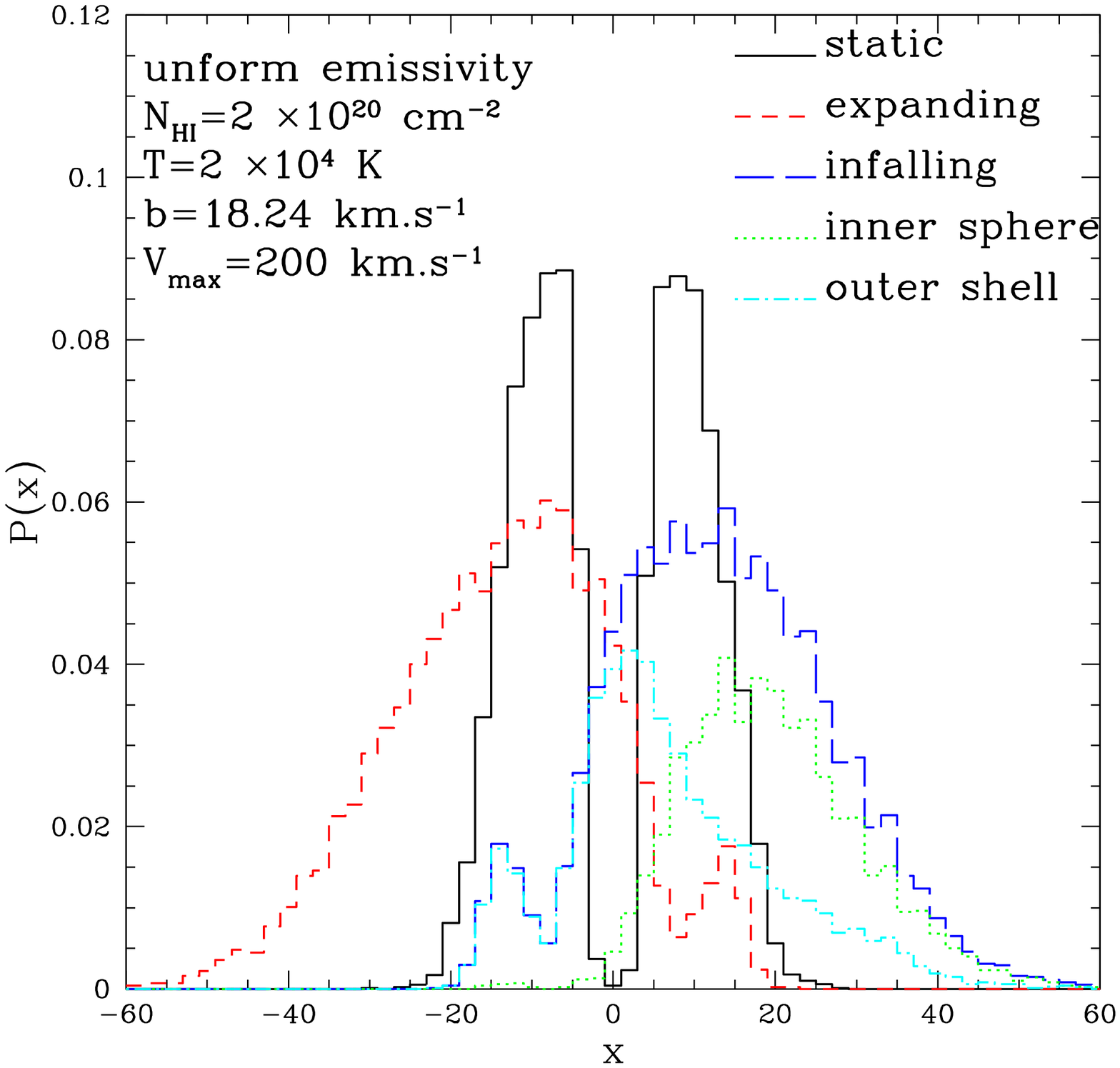} \\ 
\end{tabular}
\caption{Predicted \lya\ profiles of an isotropic and monochromatic source
in uniform and isothermal spherical halos with a center to edge column density 
of $N_H=2. \, 10^{20}$ cm$^{-2}$ and $T=20000$ K (i.e.\ $b=18.24$ \kms). 
Black solid lines show the \lya\ profiles for a static halo,
red short-dashed lines the expanding case with $\vmax=200$ \kms, 
and blue long-dashed the infall case (same $\vmax$).
{\bf Left:} Central point source. {\bf Right:} Uniform emissivity.
On the right panel, the ``infalling'' spectrum has been decomposed in two
components: the emergent spectrum from an
``inner part'' (the inner 50 \% of the volume), resembling that from a halo 
with a central source, and the one from the ``external part'' akin to 
the spectrum from an expanding shell where photons do not undergo
backscatterings (cf.\ Fig.\ \ref{spec_exp_shell}).}
\label{DLA}
\end{figure*}

\begin{figure*}[htb]
\begin{tabular}{ll}
\includegraphics[height=6cm, width=8cm]{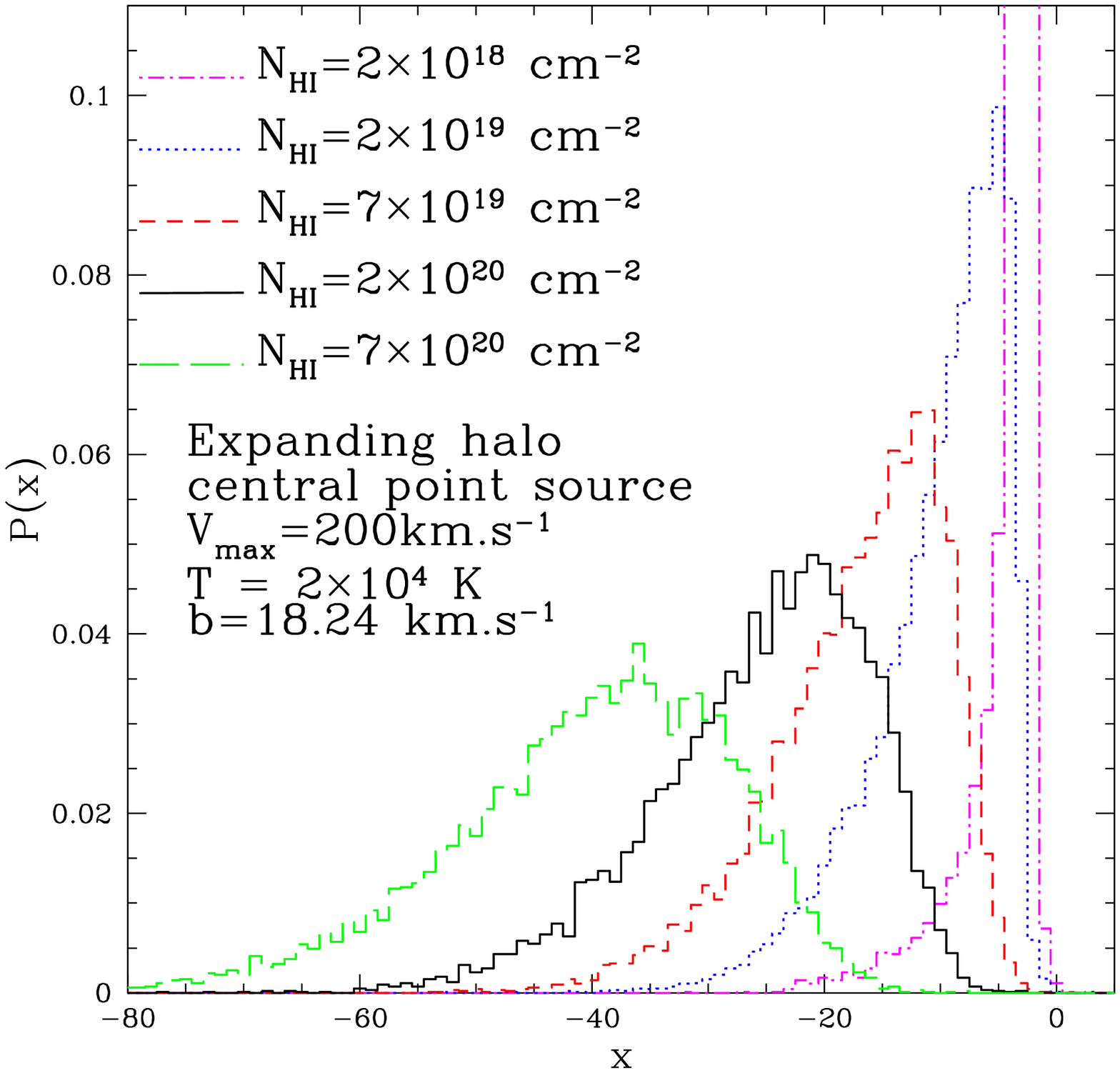} &
\includegraphics[height=6cm, width=8cm]{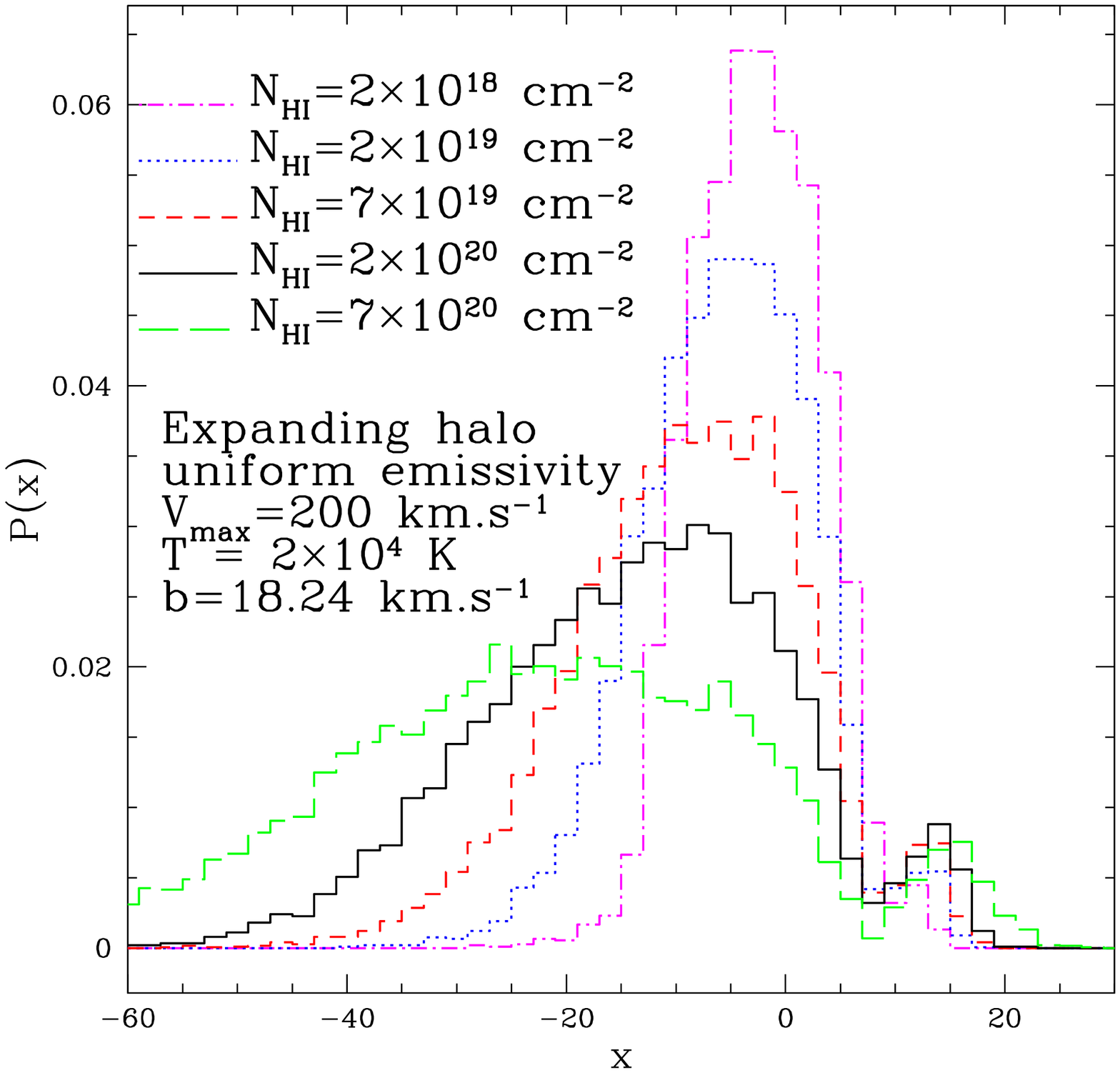} \\
\end{tabular}
\caption{Predicted \lya\ profiles of an isotropic and monochromatic source
in uniform and isothermal spherically expanding halos with 
varying column density ($N_{\rm H}= 2.\, 10^{18}$ to $7.\, 10^{20}$ cm$^{-2}$
as indicated in the Figure), and other parameters fixed to
$T=20000$ K (i.e.\ $b=18.24$ \kms) and $\vmax=200$ \kms. 
{\bf Left:} Central point source. {\bf Right:} Uniform emissivity.}
\label{evol_vs_N_halo}
\end{figure*}

We first chose the same physical conditions as Zheng \& Miralda-Escude
(2002) and performed a run with a colunm density measured from the
center to the edge of the cloud $N_{\rm H}= 2.\, 10^{20}$ cm$^{-2}$,
corresponding to a line-center optical depth $\tau_0 = 8.3 \, 10^6$
and a temperature of $2.\, 10^4$ K.  The velocity field is a
Hubble-type flow, $V(r) = \pm \vmax r/R_{\rm max}$, with 
$\vmax=200$ \kms\ at the outer radius $R_{\rm max}$ of the halo, 
and the positive (negative) sign corresponding to the expanding
(infalling) halo.
Our results, shown in Fig. \ref{DLA}, are in good agreement 
with those of Zheng \& Miralda-Escude (2002). 
For static halos, the same characteristic double peak profile as for 
the static slab (cf.\ Fig.\ \ref{slab_spec}) are obtained both for a
uniform emissivity or for a central point source. 
As shown by Djikstra \& Haiman (2005a) the position of the peaks shows
the same dependence as for the static slab (Eq.\ \ref{e_peak}), but is
given by $x_p = \pm 1.06 (a \tau_0)^{1/3}$ for $a\tau_0 > 10^3$. 

\lya\ line profiles from an expanding or infalling halo are perfectly
symmetric to each other 
(compare the red short-dashed and blue long-dashed curves on Fig.\ \ref{DLA}).
Expanding halos present a red peak, whereas infalling
halos have a blue one. Halos with a uniform emissivity  (right panel
in Fig.\ \ref{DLA}) show 
{\em 1)} broader lines, 
{\em 2)} emission extending on both sides, and 
{\em 3)} a secondary peak on the blue (red) side for the expanding
(infalling) halo. 
The last feature is not visible on the plots of Zheng \&
Miralda-Escude (2002),  but their resolution may be too low. However,
the results of Dijkstra \& Haiman (2005a) show this secondary peak. We
now briefly discuss the origin of these features. 

Why are single peaks formed in expanding/infalling media with a
central point source emitting monochromatic radiation at the \lya\
line center? The reason is simple. The probability to escape the medium for a
photon at line center is $e^{-\tau_0}$, i.e.\ close to zero for
both cases shown here. As an expanding halo contains atoms
with velocities $v(r)$ from 0 to $\vmax$, all
photons in the frequency range $x=[0,\vmax/b \sim 11]$
will be seen \emph{in the line center} by atoms of the
corresponding velocity, and are thus ``blocked''. Therefore the only 
possibility to escape is to be shifted to the red side. 
The symmetry of the double peak profile of the static case is ``broken''
in this way and transformed to a red peak for an expanding halo. 

The increased line width in case of a spatially uniform emissivity (1) and
the presence of photons on both the red and blue sides (2) is
due to the fact that the intrinsic line emission (assumed at $x=0$
in the atom's frame) spreads already over all frequencies from $x=-\vmax/b$ 
to $+\vmax/b$ in the external (observer's) restframe. 
Radiation transfer effects further redistribute the photons in the wings.
In fact, for the expanding halo, photons emerging with $x > \vmax/b$ 
(i.e.\ very blue ones) correspond to photons emitted close to the halo edge 
on the approaching side, which have been further redshifted by diffusion
away from the line center (at $x\sim +\vmax/b$ in the observers frame). 
This naturally also produces the local minimum observed at this
frequency, which separates the secondary peak from the main one (point
3 above). This is easily verified by plotting e.g.\ the contributions
from photons emitted in the ``external'' parts (i.e.\ close to the
edge) and from ``internal''regions, as shown in the right panel of
Fig.\ \ref{DLA}. We chose as inner/outer limit the radius which
corresponds to an inner sphere and an outer shell of same volume.  
Therefore the location of this minimum is a measure of the external
velocity of the halo, as already noticed by Dijkstra \& Haiman
(2005a), at least  for velocities $\vmax \la$ 360 \kms\ beyond
which the two peaks start to overlap.

\begin{figure}[tb]
\includegraphics[height=6cm, width=8cm]{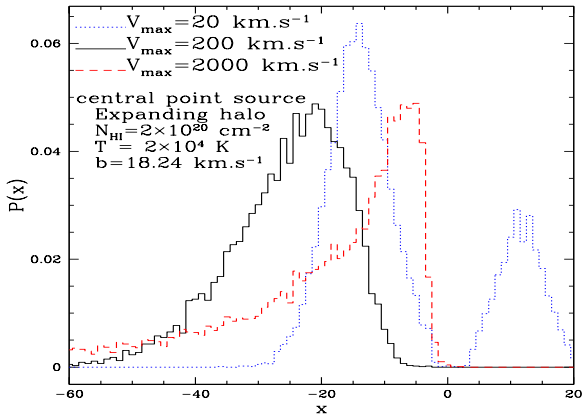} 
\caption{Predicted \lya\ profiles from an expanding halo with a
  central monochromatic source and varying velocity gradients
  ($\vmax=$ 20, 200, and 2000 \kms) for  $T=20000$ K (i.e.\ $b=18.24$
  \kms)  and $N_{\rm H}= 2.\, 10^{20}$ cm$^{-2}$.} 
\label{evol_vs_V_halo}
\end{figure}

With Figs.\ \ref{evol_vs_N_halo} and \ref{evol_vs_V_halo}, we
investigate how the emergent lines depend on column density and
velocity gradient. When $N_{\rm H}$ increases, the emission peak is shifted
away from line center and broadens out for both cases (central point source and
uniform emissivity). Indeed, the optical depth at line center increases with
$N_{\rm H}$, so \lya\ photons have to diffuse far in the wings to
escape. From the right panel of Fig.\ \ref{evol_vs_N_halo} we note
that the location of the flux minimum between the two peaks remains 
constant quite independently of $N_{\rm H}$.
The behaviour for different velocity gradients, first
 discussed by Wehrse \& Peraiah (1979),is the following
(cf.\ Fig.\ \ref{evol_vs_V_halo}):
from $\vmax=20$ to 200 \kms\ one goes from a nearly static case
(i.e.\ double peaks) to a broad asymmetric line, whose peak position
is progressively displaced redwards. Above a certain value of the velocity
gradient (or equivalently $\vmax$), the peak position moves back
closer to line center as the medium becomes more ``transparent'' there, 
given the finite and constant column density. In any case, the extent of
the red wing continues to increase with increasing $\vmax$.
Qualitatively the simulations shown in Figs.\  \ref{evol_vs_N_halo} and 
\ref{evol_vs_V_halo} resemble those of a cosmological Hubble flow
(cf.\ Loeb \& Rybicky 1999) although modified by a finite outer boundary.

For test purposes we have also compared the surface brightness profiles with 
the results of Zheng \& Miralda-Escude (2002), and find a good agreement.

\begin{figure*}[htb]
\begin{tabular}{ll}
\includegraphics[height=6cm, width=8cm]{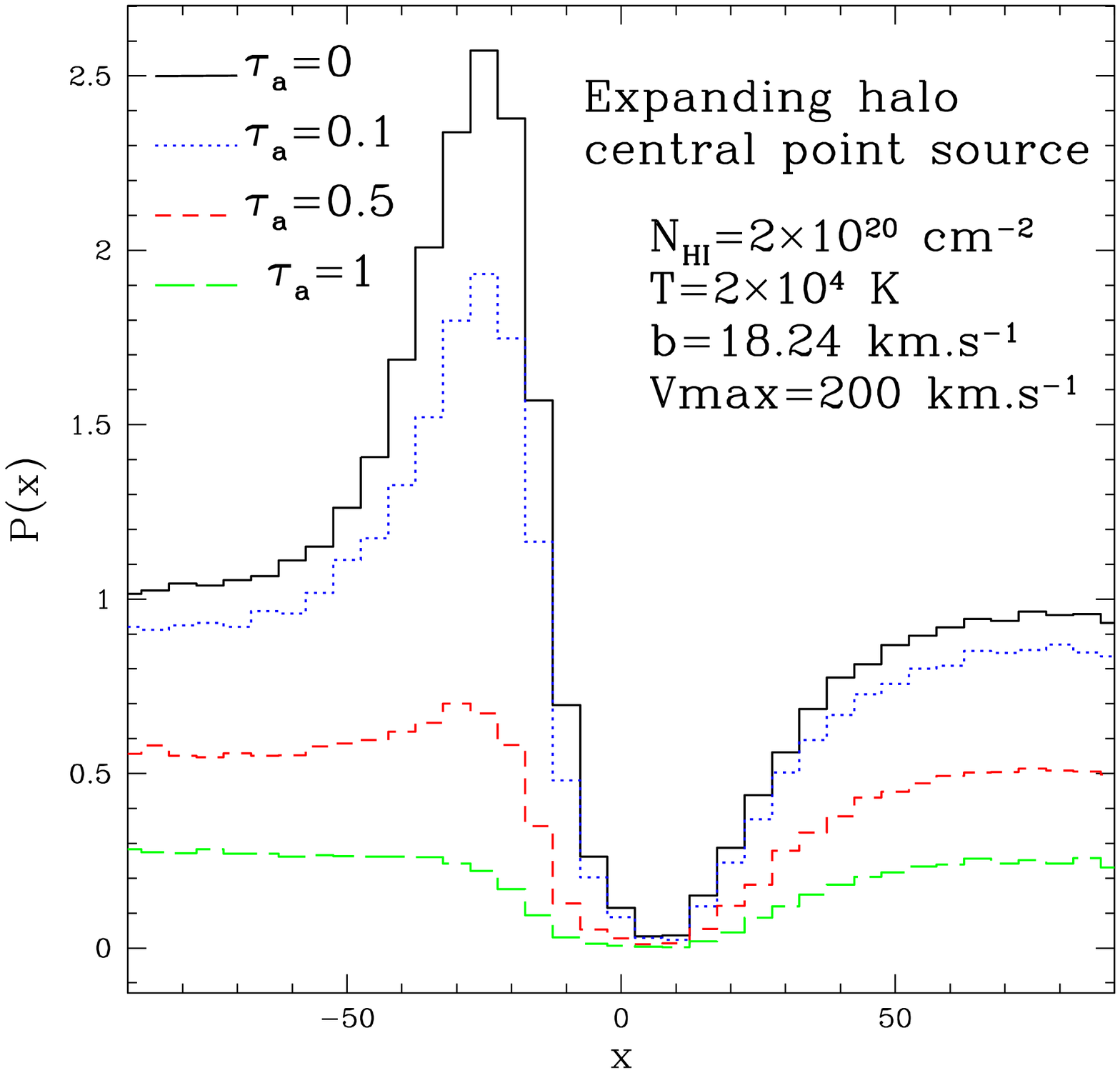} &
\includegraphics[height=6cm, width=8cm]{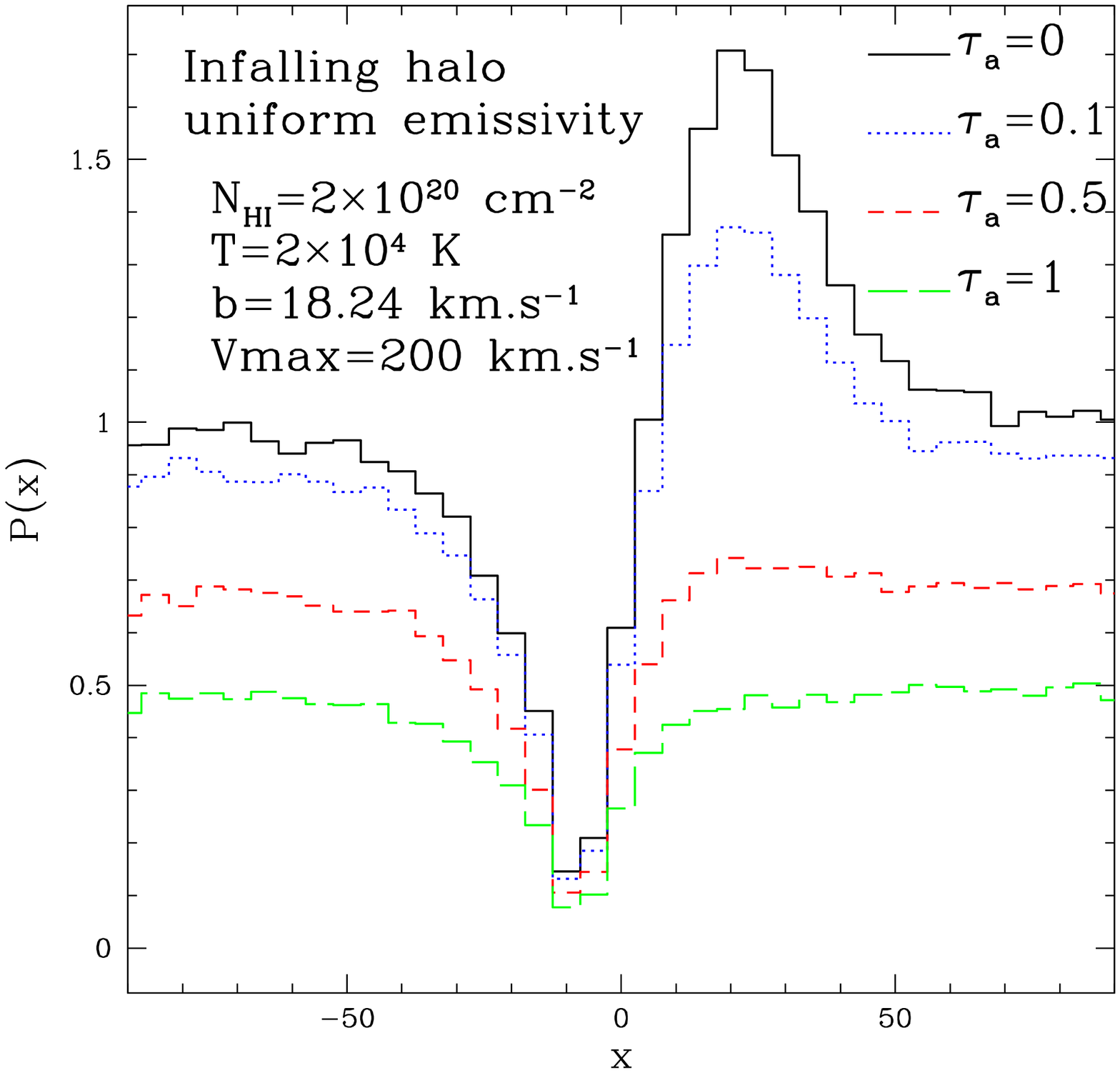} \\
\end{tabular}
\caption{{\bf Left:} Line profiles for an expanding halo with a
  central monochromatic source showing the variety of profiles from
  P-Cygni to somewhat asymmetric damped profiles.   
{\bf Right:} Infalling halo with a uniformly distributed emissivity
showing profiles from inverted P-Cygni to less broad absorption.
In both panels the adopted conditions are the following: 
$N_H=2. \, 10^{20}$ cm$^{-2}$, $T=20000$ K, and $\vmax=200$ \kms.
The line profiles are shown for dust amounts corresponding to:
($M_d/M_H$, $\tau_a$) = (0., 0.), ($8.\, 10^{-4}$, 0.1),
($4.\, 10^{-3}$, 0.5), ($8.\, 10^{-3}$, 1.).}
\label{dust_in_halos}
\end{figure*}

\subsubsection{Continuous input spectrum plus a line and dust}
\label{s_dust_in_halos}

Considering a source spectrum with a pure continuum and varying
amounts of dust yields the following for the same conditions
considered before(see Fig.\ \ref{dust_in_halos}).
Due to the scattering nature of \lya, the continuum photons are removed from 
line center and redistributed in the wings leading to a P-Cygni type profile 
with a red emission and blue absorption in the expanding medium, and
an inverted P-Cygni profile in the infalling case. In other words, a
flat source spectrum can result in a complex line profile.

The peak is located at the same frequency as in the
monochromatic case with identical physical conditions, 
and the absorption feature is around $x=\vmax/b$. 
Again its location indicates the external velocity of the medium.
The emission peak is broader for uniform emissivity than for a central
point source, as already discussed in the monochromatic case,
whereas the absorption feature behaves inversely.
The latter behaviour is due to a partial ``replenishment'' of the absorption
feature by the ``secondary peak'' discussed above for the monochromatic
radiation.
Also shown in Fig.\ \ref{dust_in_halos} is how
the presence of dust changes the observed line profile: 
as dust absorption is efficient in the wings close to (but not at) line center
(cf.\ Sect.\ \ref{s_slab_dust}) it easily suppresses the emission peak
and broadens the absorption feature leading to an asymmetric absorption
profile with a more strongly damped red (blue) wing for an infalling
(expanding) halo.

Far from the line center, the escape fraction, i.e.\ the ratio between
the number of photons which escape the medium in one frequency bin and
the number of photons emitted in this frequency bin, follows the
expected exponential law
$f_e = e^{-\tau_a},$
where $\tau_a$ is the dust absorption optical depth. Indeed, when the
influence of hydrogen scattering becomes negligible, photons are be
absorbed by dust with the probability $P=1-e^{-\tau_a}$.

\begin{figure}[tb]
\includegraphics[height=6cm, width=8cm]{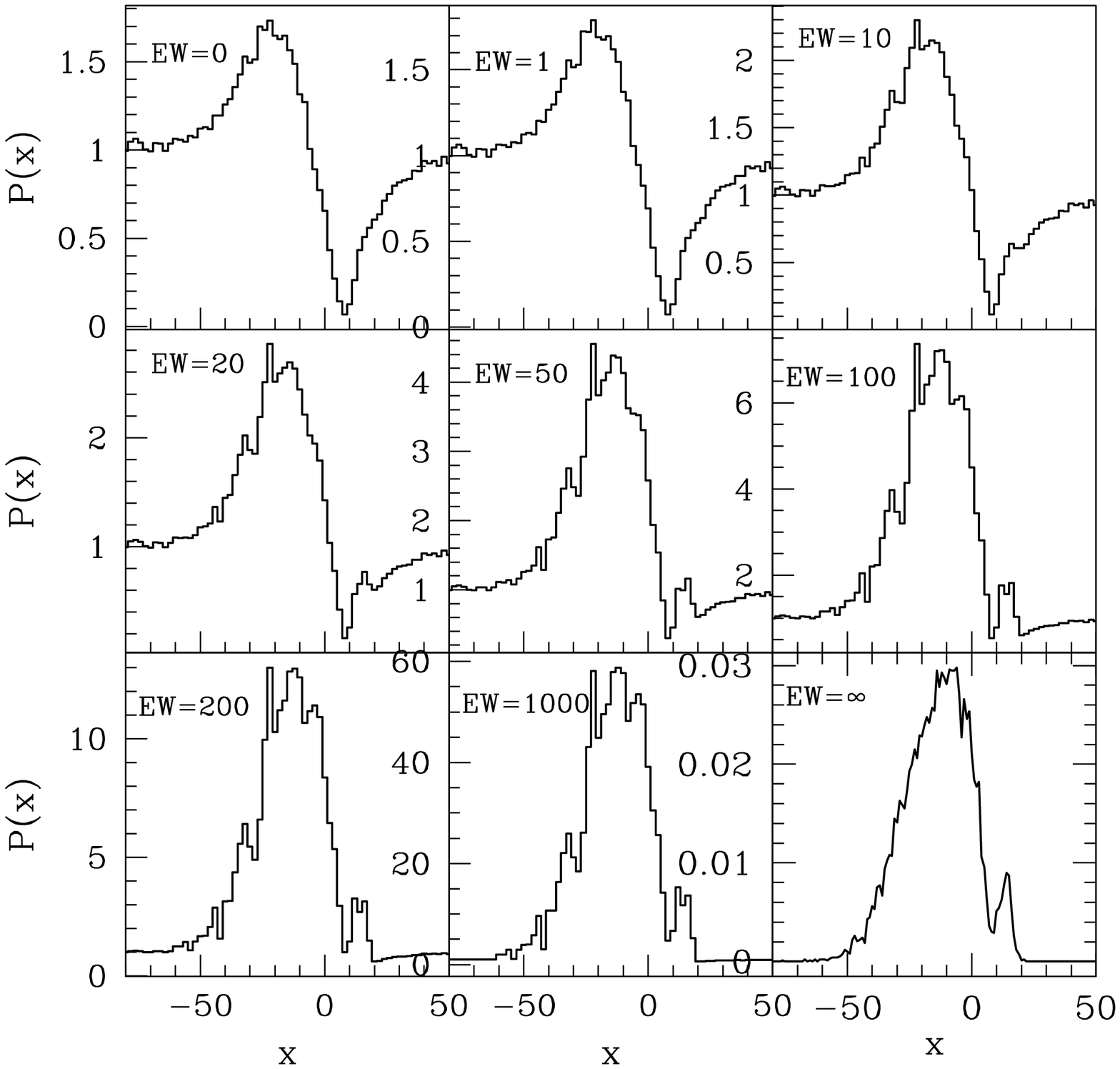}
\caption{Predicted \lya\ line profiles from an expanding halo with 
uniformly distributed sources emitting a flat continuum plus 
a monochromatic line (at \lya\ line center) with different equivalent widths.
The physical conditions in the halo are: $N_{H}=2. \, 10^{20}$ cm$^{-2}$,
$T=2\times 10^4 K$, $\vmax=200$ \kms, and no dust.
The \lya\ line equivalent width, given in units of the Doppler width $\Delta\nu_D$,
increases from $0$ to $\infty$ from 
the top-left corner to the bottom-right.}
\label{movie_halo}
\end{figure}

To illustrate the variety of line profiles predicted for various intrinsic
line strengths, we show in Fig.\ \ref{movie_halo} the case of an
expanding halo with uniform emissivity. 
As seen from this figure, a family of line profiles with intermediate cases
between a pure continuum (cf.\ Fig.\ \ref{dust_in_halos}) and pure
line emission (Fig.\ 
\ref{DLA}) is obtained with
increasing intrinsic \lya\ equivalent width $W(\lya)$.
Note that for a sufficiently large  $W(\lya)$ the secondary
peak becomes again visible. 
Also, remember that for the case of opposite movement, i.e.\ infalling,
the predicted spectrum is identical except for 
a change of red and blue frequencies (i.e.\ change $x$ to $-x$).

Overall, as we have seen, there are several degeneracies which
make it difficult if not impossible to determine physical parameters 
such as \nh\ and \vmax\ for cases of expanding or infalling halos.
This is in particular complicated by the lack of a priori knowledge 
of the spatial distribution of the emissivity and of the precise
velocity field.
For a sufficiently extended \lya\ emissivity it may be possible to determine 
the outer velocity, \vmax, from a local minimum of the flux in the blue 
part of the line profile, as already pointed out by Dijkstra \& Haiman (2005a).
However, detecting this feature would require a fairly high signal to noise.
Furthermore this blue side of the \lya\ profile may be altered
by intervening IGM absorption components at slightly lower redshifts.

\section{On the formation of damped \lya\ absorption profiles}
\label{s_slab}

As damped \lya\ profiles are frequently observed and pure Voigt absorption
line profiles often used to fit components of \lya\ it is of interest
to examine under which conditions actually pure (damped or non-damped)
\lya\ absorption line profiles are expected.

\begin{figure}[tb]
\includegraphics[height=6cm, width=8cm]{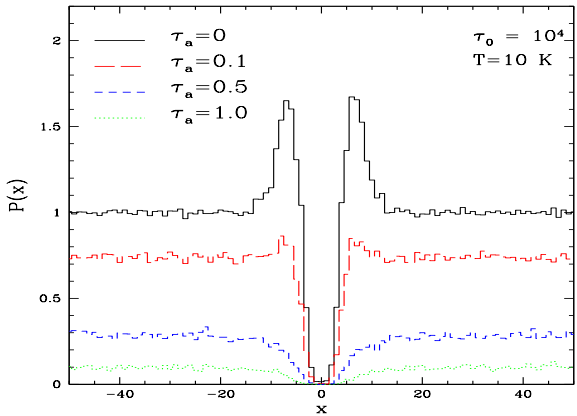}
\caption{Emergent \lya\ profiles for varying amounts of dust
predicted for a slab with an isotropic source in the center emitting 
a flat continuum. The conditions of the medium are:
$\tau_0 = 10^4$, $T=10$ K.
In this static configuration, the two angular redistributions are
  equivalent; spectra have been obtained with the isotropic redistribution.
The line profiles are shown for absorption dust optical depths of
$\tau_{\rm a}=0.$ (no dust), 0.1, 0.5, 1. from top to bottom.
Note the double-peak profile plus central absorption
characteristic of a static medium for no dust. Already small amounts
of dust are able to suppress the peaks, thereby leading to Voigt-like
profiles. 
However, in such geometrical configurations, classical Voigt-profile
fitting will underestimate the true HI column density. See text for 
more explanations.}
\label{slab_cont}
\end{figure}

To illustrate the point, and the non-trivial problem of \lya\ radiation
transfer effects we show in Fig.\ \ref{slab_cont} the predicted 
spectrum around \lya\ for an infinite slab illuminated uniformly 
in the central plane by a pure continuum source. 
Indeed from the discussion of the previous cases it is not surprising
to find a double peaked emission profile with a deep central absorption,
as shown for the case without dust.
Adding already quite small amounts of dust allows one to destroy the 
emission peaks and to create \emph{Voigt-like} absorption profiles, as also shown
on Fig.\ \ref{slab_cont}.
All these profiles can be seen as the sum of two components:
  first photons which have not undergone any interaction in the slab
  leading to a Voigt profile, added to the usual double-peaked
  profile arising from photons which scattered in the slab. The latter
  one becomes less important when the amount of dust increases
(for a fixed \nh) and/or when \nh\ increases (for a fixed dust-to-gas ratio).
Hence the resulting profile approaches a Voigt profile.
However, one should notice that the column density derived from Voigt-profile 
fitting to these predicted profiles 
can be several times smaller than the ``true'' column density in the
simulation. For example, 
for the most dusty simulation (green
long-dashed curve) from Fig.\ \ref{slab_cont} the Voigt profile yields a good 
fit (in the sense of a reduced $\chi^2 \sim 1.02$), and the best fit
column density is a factor $\sim 3.1$ lower than the ``true'' input value.

The reason for this apparently ``strange'' behaviour is simply due to 
the fact that the predicted spectrum is computed here from all the emergent 
photons integrated over the whole object and over all emergent directions,
i.e.\ it corresponds to an integrated spectrum of a symmetric object (e.g.\
the sphere with a central source) or of a sufficiently extended ``screen'' 
between us and the source.
In this case we include all photons, some of which have undergone a 
complex scattering history before emerging toward the observer.
Therefore the double-peak, characteristic of static media is unavoidable
without dust and the only way to ``destroy'' them is by adding dust.

Considering different geometries such as a finite slab, or a sphere
of the same center-to-edge optical depth and same temperature yields
exactly the same line profiles as the infinite slab.

\begin{figure}[tb]
\includegraphics[height=6cm, width=8cm]{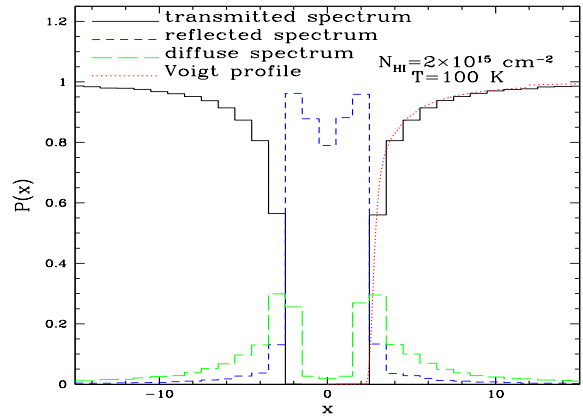}
\caption{Predicted \lya\ profiles of the direct transmitted (black
  solid line),  the reflected (blue short-dashed), and the diffuse
  (green dotted) radiation from a static slab illuminated
  perpendicularly by a distant background source with a flat continuum.
The properties of the intervening slab are:
$N_{\rm H} = 2. \, 10^{15}$  cm$^{-2}$, $T=100$ K (i.e.\ $b = 1.29$ \kms).
The directly, i.e.\ perpendicularly, transmitted flux yields the expected
Voigt profile (red dotted line). The diffuse (scattered) light 
shows again the double peak profile as the slab with a central source.}
\label{slab_cont_ext}
\end{figure}

The ``implicit'' geometrical assumptions made above for the construction of
an integrated spectrum are unlikely to be applicable to the typical 
damped \lya\ systems (DLAs) observed in spectra of distant galaxies and
quasars. These cases are better idealised by a small (in angular size)
intervening cloud  between the background source and the observer, 
which diffuses basically all photons out of the line of sight.
In other words photons far from the line center will simply travel
freely through the cloud, but when the photon frequency $x$ is close
to the line center, the probability to cross the medium without
interaction, $\exp(-\tau_0 H(x,a))$, is considerably reduced.
Since photons which interact with the medium have basically no chance
to be re-emitted along the same line of sight (compare the observer
solid angle with $4\pi$), they will be lost for the
observer. Therefore the observed profile is simply that of a pure
absorption line described by the Voigt absorption line profile, namely
$I(x) = I_0 \exp(-\tau_0 \, H(x,a))$, reflecting the properties of the
medium, i.e.\ the total column density and the temperature.

We simulated this configuration, and
the resulting spectrum is presented on Fig~\ref{slab_cont_ext}, by the
black curve.
We checked again that a spheric cloud or a slab perpendicular to the line 
of sight, with the same optical depth and temperature, lead exactly to the 
same observed spectrum. For reasons of calculation time, we considered a 
relatively small column density, $\nh = 2 \times 10^{15} cm^{-2}$. 
The temperature is set to $T = 100 K$.
As expected, the simulation is perfectly fitted by a theoretical Voigt
profile (red dotted curve).
Also plotted on the same graph is the reflected spectrum (blue curve), 
composed of photons which escaped the medium by the side they entered. 

Photons which are not ``reflected'', i.e.\ backscattered to the source will diffuse
in the medium and finally escape after a large number of
scatterings forming the ``diffuse spectrum''. For obvious reasons 
this spectrum, shown by the green curve, presents the same
shape as the emergent spectrum from a slab with an isotropic source in
the center, i.e.\ two symmetrical peaks. 
In dust-free cases with a flat incident continuum source, the transmitted 
spectrum will be the opposite of the reflected + diffuse spectrum, as
all \lya\ photons are conserved.

\begin{figure*}[htb]
\begin{tabular}{ll}
\includegraphics[height=6cm, width=8cm]{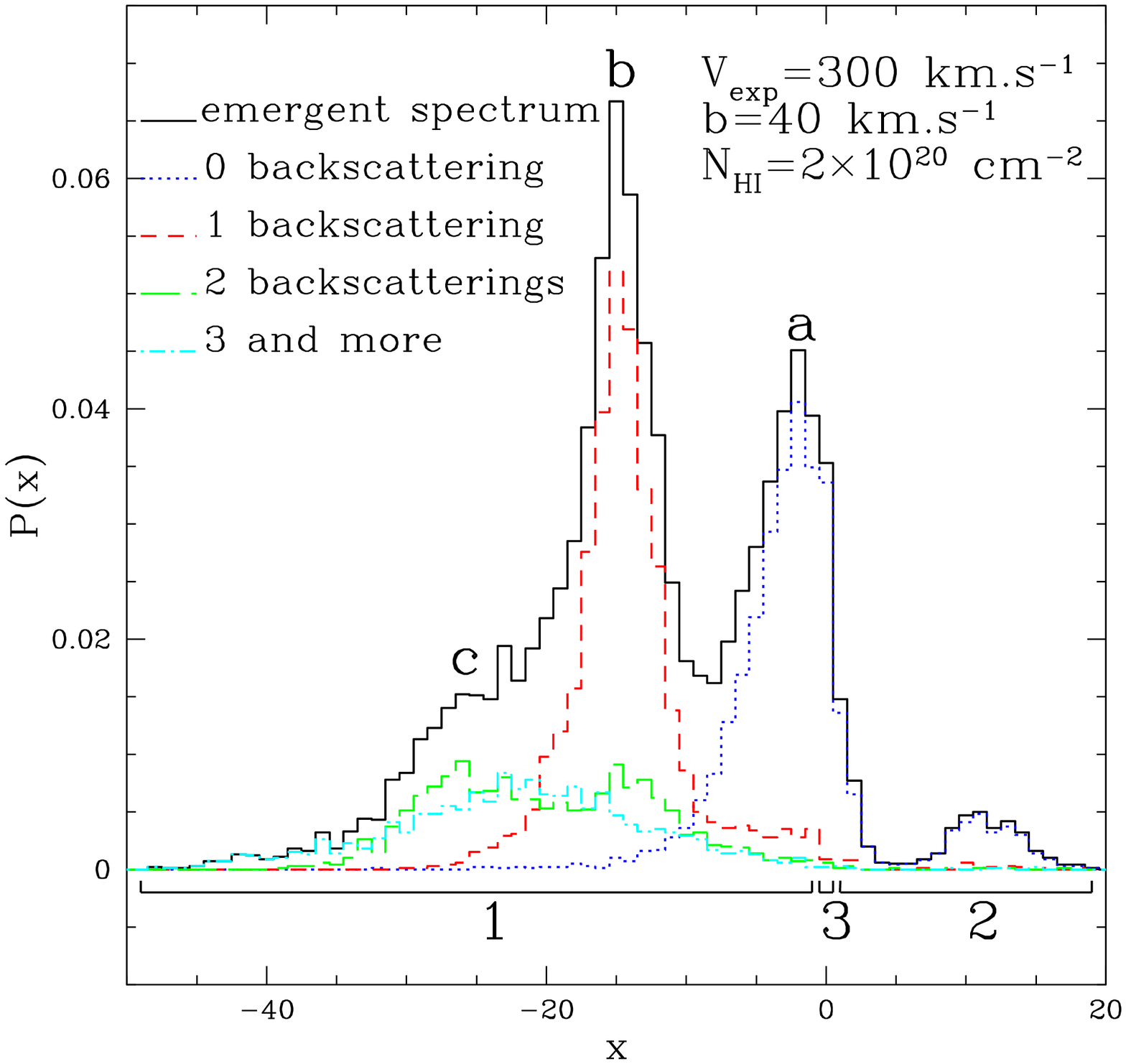} &
\includegraphics [height=6cm, width=8cm]{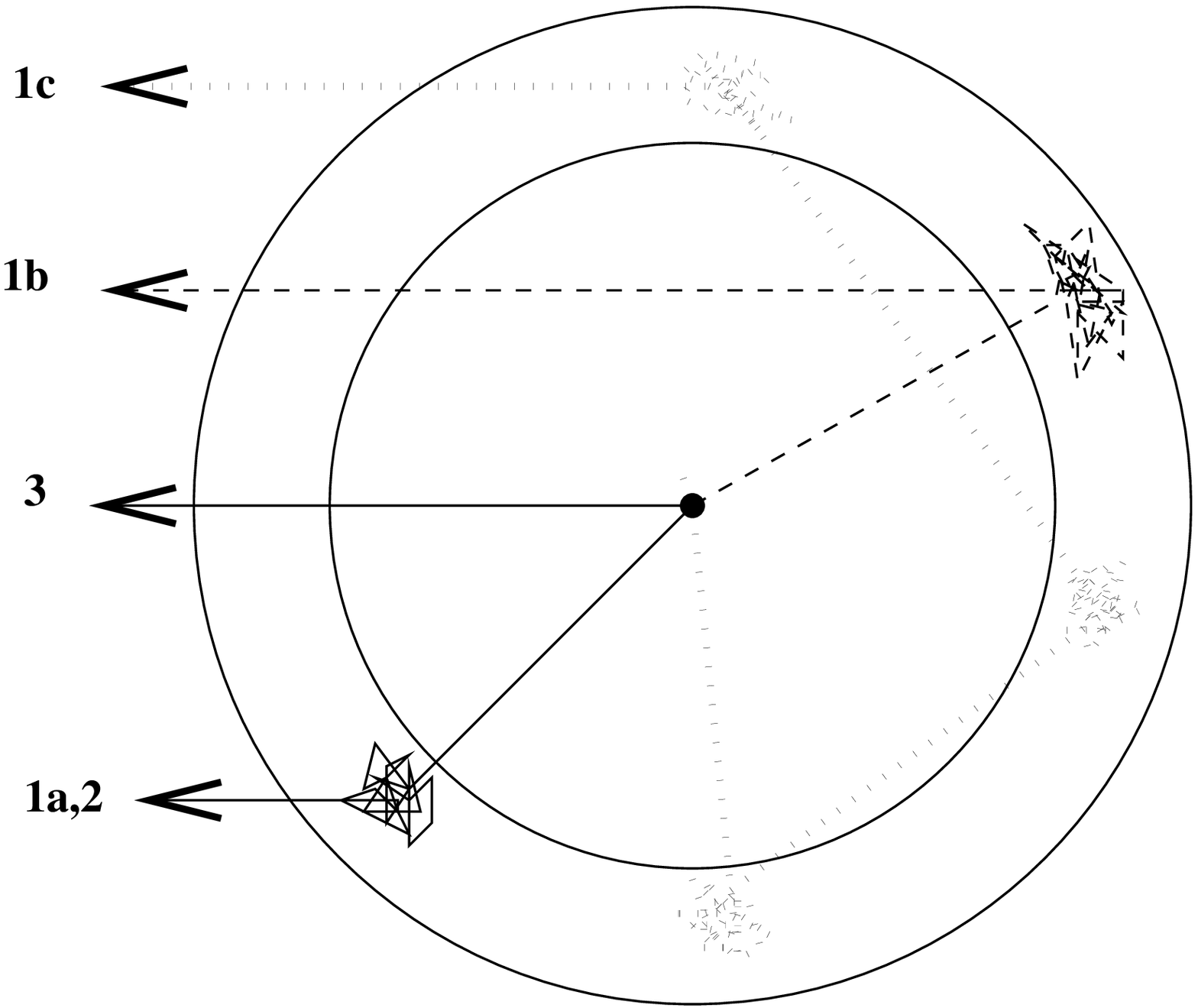} \\
\end{tabular}
\caption{{\bf Left:} Emergent \lya\ profile from an expanding shell
with central monochromatic source. The different
  shapes can be described with the number of back-scatterings photons
  undergo : bumps $1a$ and $2$ are built-up with photons which didn't
  undergo any backscattering, the highest peak located at $x= -2
  \vexp/b$ (feature $1b$) is composed of photons which undergo
  exactly one backscattering, and the red tail $1c$ is made of photons
  which undergo two or more backscatterings. See text for more details.
{\bf Right:} Scheme of an expanding shell showing the origin of photons
  of different groups. The observer is on the left side, at
  infinity.}
\label{spec_exp_shell}
\end{figure*}

In short, to form pure symmetric \lya\ absorption line profiles from 
a flat continuum source requires specific geometrical configurations
allowing the photons to diffuse out of the line of sight. 
An alternative way to achieve such profiles, e.g.\ in an integrated spectrum
of an embedded source, is by invoking the presence of dust which
destroys the double emission peaks otherwise present. 
However, in such a case the apparent column density derived from simple Voigt profile
fitting underestimates the true value of \nh\ due to radiation transfer
effects.

\section{\lya\ transfer in expanding dust-free and dusty shells revisited}
\label{s_shell}

There are numerous indications, both theoretically and observationally, 
for the presence of expanding shells and bubbles in starbursts.
It is therefore important to simulate such geometrical configurations 
to examine both qualitatively and quantitatively the diversity
of \lya\ line profiles and to gain basic insight into the physical
processes shaping them.

Our model of an expanding homogeneous shell is described by the following
parameters:
an inner and outer radius $R_{\rm min}$ and $R_{\rm  max}$ respectively, 
a uniform radial expansion velocity \vexp, the radial colunm density \nh, 
and a constant temperature $T$ given by the Doppler $b$ parameter
(Eq.\ \ref{e_b}). The interior of the shell is assumed to be empty,
the isotropic \lya/continuum emitting source located at the center.
In contrast to the geometrical configurations discussed earlier,
  emergent profiles from expanding shells are sensitive to the angular
  redistribution.
Therefore the dipolar redistribution is taken into account to treat 
the \lya\ radiation transfer consistently in this case, and
all spectra shown in this section have been obtained using
the dipolar redistribution in the code.
The main parameters determining the \lya\ photon escape and hence the
line profile are \vexp, \nh, $b$, and the dust-to-gas ratio $M_{\rm
  d}/M_{\rm H}$, as shown below. The precise values of the shell radii
and thickness, setting the geometrical size and curvature, are secondary.

Below we shall examine the following cases:
The academic cases of a source with monochromatic emission and shells
both without and with dust. These first two cases are essential to
understand more realistic simulations allowing for arbitrary input
spectra (including continuum and/or \lya\ line emission), again
dust-free or with dust, discussed subsequently.

\subsection{Monochromatic emission and dust-free shells}

\subsubsection{Basic line profile formation mechanism}

First we discuss the emergent \lya\ spectrum from a dust-free expanding shell
with the following parameters:
$\vexp=300$ \kms, $\nh=2\times 10^{20}\,cm^{-2}$, $b=40$ \kms\ (i.e.\
$\tau_0 = 3.8\times 10^6$),  $R_{\rm min} = 0.9 R_{\rm max}$, and 
$R_{\rm max}=1.17\times 10^{19}$ cm.
For monochromatic \lya\ photons emitted at line center the resulting 
line profile is shown on the left panel of Fig~\ref{spec_exp_shell}
(solid black line). Qualitatively this line profile and others shown
below exhibit the following characteristic features, marked on the figure and
illustrated on the right panel:
\begin{itemize}
\item[] {\bf 1abc)} an extended redshifted emission with one or two
  ``bumps'' (1a and 1b) and a red tail (1c) -- all at $x<0$ (i.e.\ red
  side of \lya),  
\item[] {\bf 2)} a smaller blue bump (at $x>0$), and 
\item[]{\bf 3)} an narrow emission peak at the line center ($x=0$).
\end{itemize}
Although we considered the same physical conditions as Ahn \etal\ 
(2003, their Fig.\ 2), features 2) and 3) are not apparent in their
simulations (cf.\ also Ahn 2004).
However, these features are also found in the simulations of Hansen \&
Oh (2006) and the origin of all of them is well understood, as we
shall now discuss.  
To do so it is instructive to group the emergent photons 
and to distinguish the emergent line profiles according to the number 
backscatterings they have undergone (see left panel on Fig.\
\ref{spec_exp_shell}).  A photon is said to ``backscatter''
when it travels across the empty interior before re-entering the shell
at a different location\footnote{This definition is equivalent to the one
used by Ahn \etal\ (2003).} (see right panel of Fig.\ \ref{spec_exp_shell}).
Note, that any such travel is counted as a backscattering, irrespective of 
its precise direction/length. In particular this does not necessarily 
imply a hemisphere change for the photon.

{\bf Features 1a and 2: photons with zero backscattering.} 
In this simulation all photons are emitted at line center ($x=0$) at 
the center of the shell. Once a photon reaches the shell for the first time
it is seen redshifted to $- x_{\rm CMF}(\vexp) = -7.5$ by the H atoms
(in the comoving frame, CMF).
A fraction of the photons will diffuse progressively through the shell towards
the exterior and escape without backscattering (solid lines on the
left panel of Fig.\ \ref{spec_exp_shell}). Their spectrum (marked as 
0 backscattering) gives rise to an asymmetric double peak with
a small blue component centered at $x_2 \sim 10$ (feature 2 above) from 
photons escaping the blue wing of \lya\ in the (blueshifted) shell
approaching  the observer, and
a somewhat redshifted stronger peak at $x_{1a} \sim -5$ (bump 1a above)  
corresponding to the photons escaping the red \lya\ wing in the
blueshifted shell. Qualitatively this part of the spectrum is
equivalent to the spectrum of a slab with a constant receding
macroscopic velocity with respect to the emitting source (see Neufeld
1990, Fig.\ 6). 

\begin{figure}[tb]
\includegraphics[height=6cm, width=8cm]{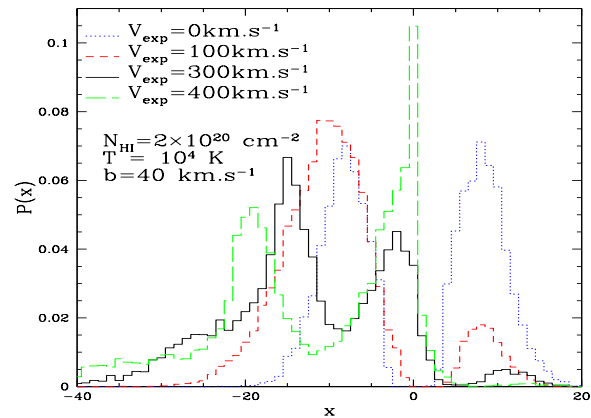}
\caption{Dependence of the predicted \lya\ profiles from an expanding shell 
with a central monochromatic source on the expansion velocity.}
\label{exp_shell_V}
\end{figure}

{\bf Feature 3: direct escape.}
For sufficiently small column densities and/or large expansion velocities,
a non-zero fraction of photons traverses directly the shell without
interacting. This case appears with the probability $e^{-\tau_{\rm
    exp}(0)}$, where $\tau_{\rm exp}(x) = \tau(x-x(\vexp)) $
is the reduced \lya\ optical depth seen by a photon with observers
frequency $x$ due to Doppler shift of \vexp.
For the case discussed here
$\tau_{\rm exp}(0)=\tau(-7.5)=6.09$, 
so that 0.25 \% of the photons will escape without interacting. 
These photons give raise to feature 3 in the bin at $x_3=0$ labeled in 
Fig.\ \ref{spec_exp_shell}.
The importance of these direct photons increases of course with increasing
\vexp\ and decreasing column density, as seen in Figs.\ \ref{exp_shell_V}
and \ref{exp_shell_N}. 
Before comparison with observed spectra this peaked flux contribution 
must obviously be convolved with the instrumental resolution. 

\begin{figure}[tb]
\includegraphics[height=6cm, width=8cm]{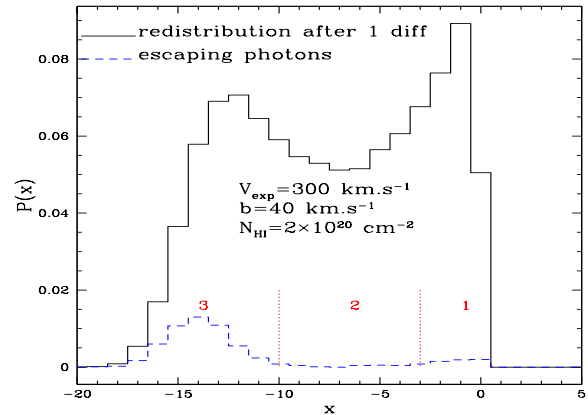}
\caption{Frequency distribution of the photons in the expanding shell
  after the first scattering. The black solid curve contains all
  photons, the blue dotted one represents the histogram of photons
  which escaped after only one scattering.  
  bf They form a bump around $x \sim -2x(\vexp)$ which explains the
  appearance of feature 1b.
 See the text for more details.}
\label{after1diff}
\end{figure}

{\bf Features 1b and 1c: photons undergoing one or more backscattering.}
Let us now examine  the situation after one scattering in the expanding shell.
After the first scattering the frequency distribution of the photons 
in the external frame is 
shown by the black solid line on Fig.~\ref{after1diff}. 
As the photons are initially seen by the atoms at the frequency
$-x(\vexp)$, i.e.\ in the wings, they are basically re-emitted at the
same frequency in the atoms frame. 
Depending on their emission direction this leads to a 
range of frequencies in the observer frame reaching from $x \sim -2
x(\vexp)$ to 0, with more photons re-emitted in the absorption
  direction (around $x \sim -2 x(\vexp)$) or in the
opposite direction ($x \sim 0$) because of the dipolar angular
redistribution (the frequency distribution for isotropic
redistribution would be nearly a square profile over $x \sim [-2
x(\vexp), 0]$). 
Which of these photons are now able to escape the medium after just one 
scattering is represented by the blue dotted histogram.

Overall, depending on their frequency  being in one of the 
3 spectral regions indicated in Fig.\ \ref{after1diff}, the fate of
the photons is as follows.

\begin{itemize}
\item {\em Frequency range 1:}  Photons with a frequency $\xobs \sim
  0$ cannot escape the medium:  
  they are emitted outward in the radial direction with a frequency too
  close to the line center (in the atoms frame). Their escape
  probability is negligible. 
\item {\em Range 2:} Around $\xobs \sim -7.5$, although their
  frequency is far from line center, no photons escape the medium for
  geometrical reasons: their emission direction is perpendicular to
  the radial direction increasing thus strongly the geometrical path
  and the optical depth.  
\item {\em Range 3:}  Most of the photons escape with a frequency
  $\xobs \la -9$ since their frequency is very far from line center,
  and their emission direction is convenient: they cross the inner
  part of the shell (backscattering), and when they arrive on the
  other side, the combination of their frequency and direction with the local
  macroscopic motion favours their escape. 
  The frequency distribution function
  after one scattering shows that the number of photons re-emitted
  with a frequency $x<-2 x(\vexp)$ decreases very rapidly. Therefore 
  the photons escaping after just one scattering already show a peak close 
  to the frequency corresponding to twice the expansion velocity. So a peak
  of escaping photons centered at the frequency $x=-2 x(\vexp)$ appears.      
\end{itemize}

Photons undergoing further scatterings will be absorbed again, and
the escape of those re-emitted around $-2 x(\vexp)$ will be favoured again
for same reasons.
  This explains why the most prominent feature (1b) in the red part
is located at $x_{1b} \sim - 2 x(\vexp)$, measuring therefore twice
the shell velocity.
Photons undergoing progressively more scatterings will show a
broadening frequency distribution compared to that after one scattering.
The broadening of its red wing is responsible for the last feature
(1c) made of photons escaping after 2 or more backscatterings.

\begin{figure*}[htb]
\begin{tabular}{ll}
\includegraphics[height=6cm, width=8cm]{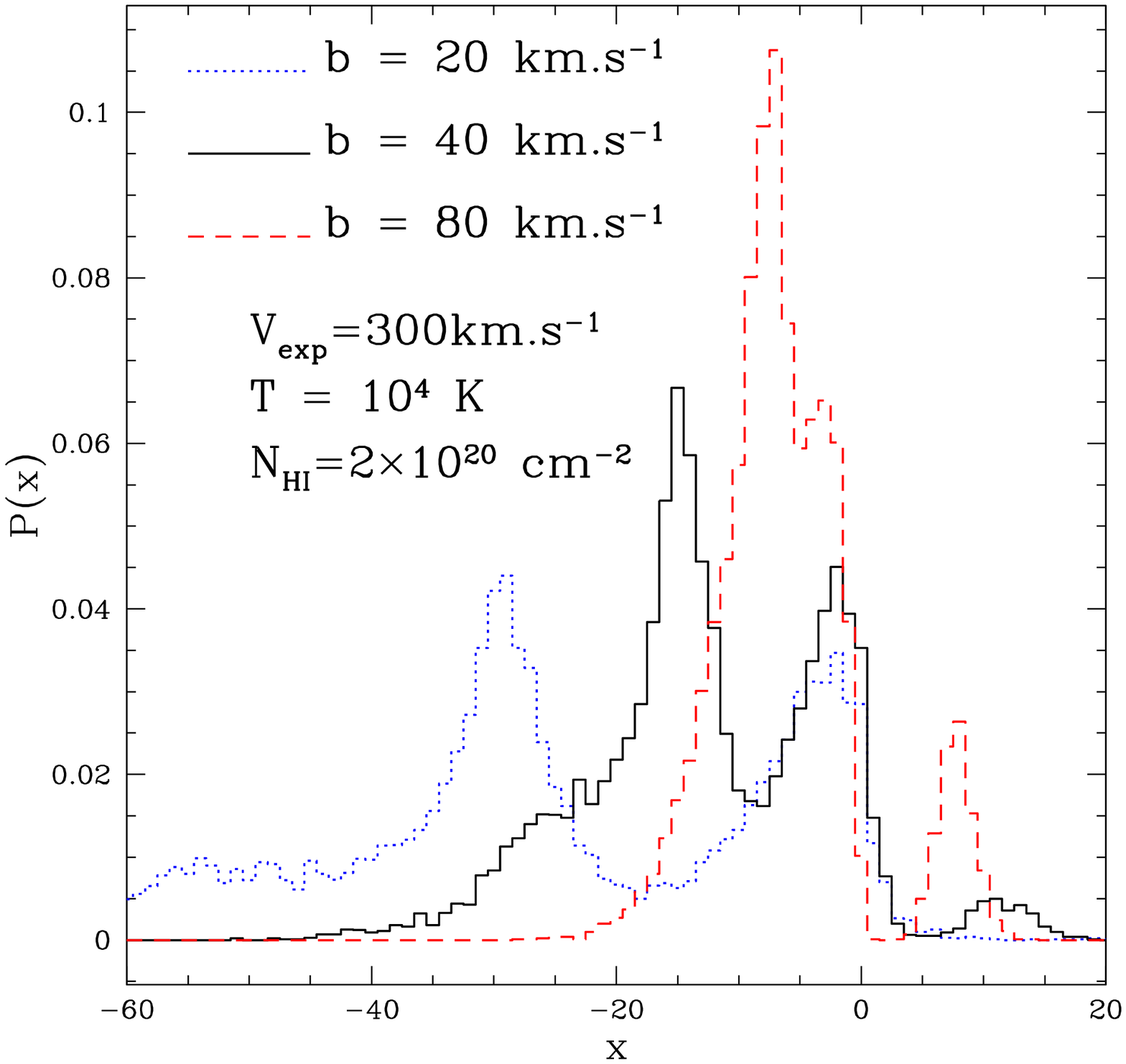} &
\includegraphics[height=6cm, width=8cm]{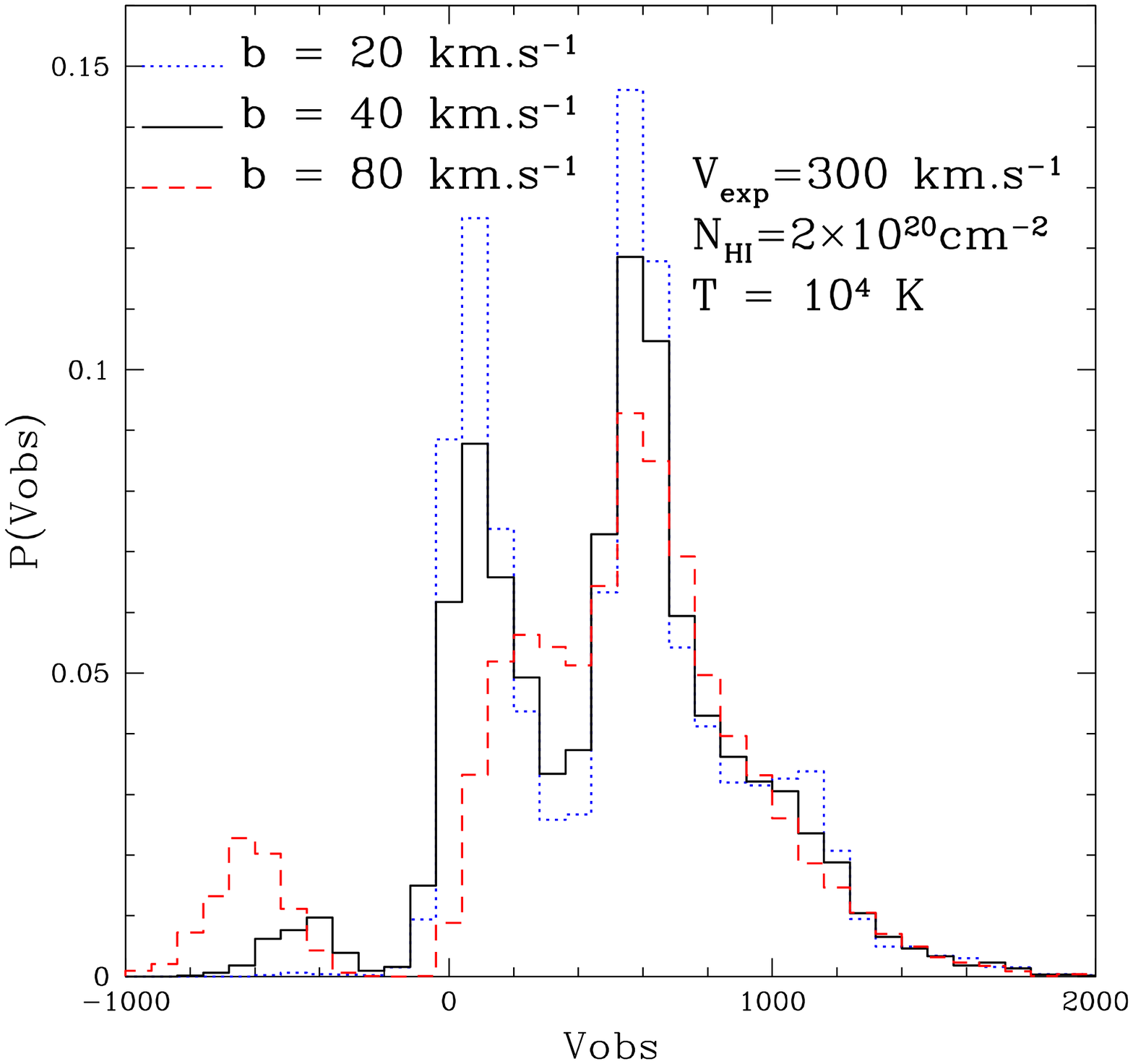} \\
\end{tabular}
\caption{Dependence of the predicted \lya\ profiles from an expanding shell 
with a central monochromatic source on the
  Doppler parameter $b$. As the characteristic unit of \lya\
  radiative transfer problem, $x$, depends on the Doppler parameter
  $b$(cf.\ Eq.~\ref{eq_x}), a variation of $b$ will lead to a change of $x$
  too. To compare cases with different $b$, we show the usual plot in
  $x$ units but also a plot in velocity. {\bf Left:} Plot in Dopper
  units $x$, {\bf Right:} Plot in velocity along the line of sight
  $v_{\rm obs}$.}
\label{exp_shell_b}
\end{figure*}

\subsubsection{Dependence on shell parameters}

Another way to understand the different features of the
emergent  \lya\ line profile of an expanding shell is by varying the
parameters. Let us examine how the spectrum depends on the
expansion velocity $\vexp$, on the thermal and turbulent velocities
intervening in the Doppler parameter $b$, and on the
colunm density $N_{\rm H}$. 

For increasing expansion velocities, and starting from the static case
with a symmetric double peak profile (cf.\ above and
Eq.\ B18 of Dijkstra \& Haiman 2005a for a static sphere),
the imbalance between the red part and the blue part grows 
(see Fig.~\ref{exp_shell_V}):
progressively more photons escape from the red part of
the line because atoms see them already redshifted (red dashed curve) at the
first interaction. The probability to be re-emitted in the line core is
then smaller than the one to ``remain'' in the wing. Hence the growing 
asymmetry between red and blue.
Note also the appearance of excess flux at line center for the curve with 
the highest $\vexp$ in Fig.~\ref{exp_shell_V}. The appearance and strength
of this feature ($x_3=0$, feature 3) is consistent with the increasing
direct escape probability. 

For a given $b$, the red part is a single peak, for small enough values 
of $\vexp$  (e.g.\ $\vexp < 200$ \kms\ for $b=40$ \kms).
In fact, the two contributions of photons undergoing zero and 
one backscattering are then too close to be distinguished, and the two
corresponding red peaks (features 1a and 1b) are mixed. 
The same trend is found for large values of $b$, as shown in
the left panel of Fig.~\ref{exp_shell_b}.
More precisely, predicted \lya\ profiles from an expanding shell present only 
one red peak for sufficiently small $\vexp/b$. 
With increasing \vexp\ or decreasing $b$,
the blue part (feature 2) becomes very faint, and almost all photons
escape in the red part, presenting two well separated peaks.
In all cases, and quite independently of $b$ (cf.\
Fig.~\ref{exp_shell_b}, left), the second and most prominent peak
(feature 1b) traces twice the expansion velocity as already discussed
above. It is essentially composed of photons which undergo one
backscattering, while the first peak (feature 1a) is made of photons
undergoing no backscattering.  
Note, however, that in contrast to the appearance in this plot, 
the position of main peak is independent of $b$ in observed spectra,
as illustrated in the right panel of Fig.~\ref{exp_shell_b}.
This is simply due to the definition of $x$, which depends on $b$.

\begin{figure}[tb]
\includegraphics[height=6cm, width=8cm]{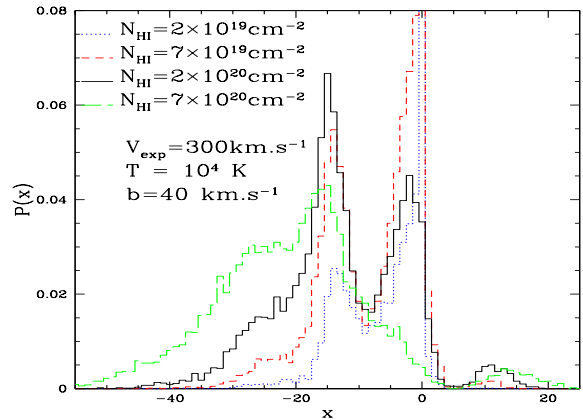} 
\caption{Dependence of the predicted \lya\ profiles from an expanding shell 
with a central monochromatic source on 
the column density.}
\label{exp_shell_N}
\end{figure}

Varying the column density (leaving the other parameters unchanged) 
has the following effects (see Fig~\ref{exp_shell_N}).
First, the flux excess at $x_3=0$, i.e.\ the fraction of photons which
can escape  the shell without interacting, decreases exponentially
with increasing $N_H$. 
Second, the relative importance of the two red peaks (1a, 1b) changes:
 with increasing column density the first peak (feature 1a) decreases with
respect to the second one (1b), whereas the red wing (1c) is enhanced,
since the importance of backscattering increases.
For sufficiently large $N_H$ 
the first peak (1a) disappears completely, whereas the red tail (1c)
becomes as important as the mean peak (1b).

In short, the imbalance between blue and red emission increases with
increasing $\vexp/b$. The separation between the multiple peaks formed
on the red side of the \lya\ profile becomes progressively less clear
(i.e.\ the peaks merge together) for lower expansion velocities,
and/or higher temperature or turbulent velocities. 
For sufficiently large \hi\ column densities (\nh\ $\ga 10^{20}$ cm$^{-2}$)
the main red emission peak measures quite well $2 \vexp$.

\subsection{Monochromatic emission and dusty shells}

\begin{figure}[tb]
\includegraphics[height=6cm, width=8cm]{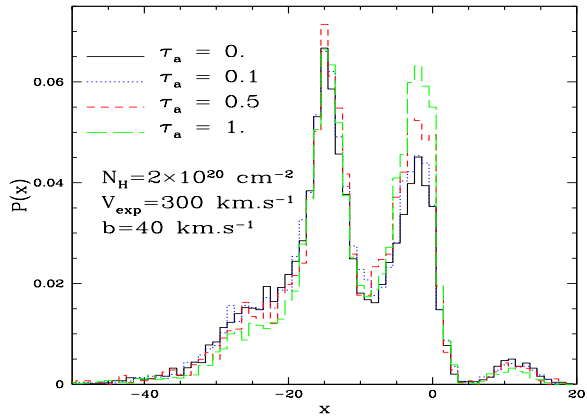}
\caption{Predicted emergent \lya\ profiles from an expanding shell with a monochromatic
central source with a column
  density $N_{\rm H}=2\times 10^{20}$ cm$^{-2}$, an expansion velocity
  $V_{exp}=300$ \kms\, a Doppler parameter $b=40$ \kms, and different dust
  absorption optical depths $\tau_a=$0., 0.1, 0.5, and 1.0. The
  area under each curve is normalised to unity, i.e.\ corrected for
  photon destruction by dust, to highlight the relatively minor line 
  profile changes. The
  blue bump (2) and the red tail (1c) are more destroyed by dust than
  other features. Overall the profile is globally preserved, but somewhat 
``sharpened''.}
\label{spec_shell_dust}
\end{figure}

In Fig.~\ref{spec_shell_dust} we present the influence of dust on the
emergent spectrum from the expanding shell studied above
(Fig~\ref{spec_exp_shell}).
 When $\tau_a$ increases, the relative height of the two red peaks
  (features 1a and 1b) is reversed: in dust-free media the prominent
  peak is 1b, but when dust is present 1a becomes as high as 1b.
In the most dusty cases, corresponding to a destruction of
$\sim$ 93\% of \lya\ photons, one notes a loss of photons from the  
the blue bump (feature 2 on Fig.~\ref{spec_exp_shell}) and from
the red tail (feature 1c on Fig~\ref{spec_exp_shell}). This is easily 
understood as these features are composed of photons undergoing a very 
large number of scattering, which increases their chance to be absorbed 
by dust.
Qualitatively our results show the same behaviour as the outflowing shell
with holes or clumps modeled by Hansen \& Oh (2006): strongly redshifted photons
are suppressed by dust, whereas the spectral peaks are still visible.
The result is a somewhat ``sharpened'' line profile.
Although this overall ``sharpening'' trend is also found in the earlier simulations
of Ahn (2004) our results differ quite strongly from theirs, as already
mentioned above. 

\begin{figure}[tb]
\includegraphics[height=6cm, width=8cm]{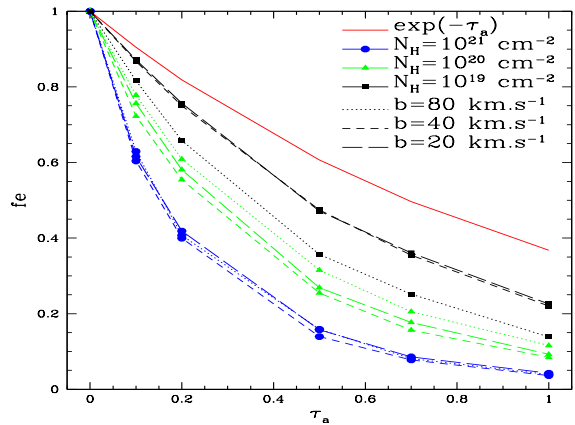}
\caption{Escape fraction of \lya\ photons from an expanding shell 
with \vexp=200 \kms for different column
  densities ($\nh=10^{19}$, 10$^{20}$, $10^{21}$ cm$^{-2}$) and for
  different Doppler parameters ($b=$20, 40 and 80 \kms).}
\label{esc_frac_exp_shell}
\end{figure}

We now quantify the \lya\ photon destruction by dust and its
dependence on the shell and dust parameters. For illustration we have
chosen similar conditions as those discussed by Ahn (2004), namely a
shell with an inner and outer radius $R_{\rm min}=0.9 R_{\rm max}$, an
expansion velocity $\vexp=200$ \kms, a \hi\ column density between
$N_{\rm H}=10^{19}$ and $10^{21}$ cm$^{-2}$, a Doppler parameter of
$b=$ 20, 40, and 80 \kms, and a monochromatic central point source.
In contrast to Ahn (2004) we assume that there is no dust inside the
bubble, i.e.\ at $R < R_{\rm min}$, as most of the dust is probably
destroyed there.  Furthermore, the presence of dust inside the shell
is not compatible with inferences from the empirical Calzetti
attenuation law (see e.g.\ Gordon, Calzetti \& Witt 2003).
The predicted \lya\ escape fraction as a function of the dust absorption
optical depth $\tau_a$ measuring, for a given \nh\ column density, 
different dust-to-gas amounts, is shown in Fig.\ \ref{esc_frac_exp_shell}.

As expected, the main dependence of the escape fraction $f_e$ is on
\nh\ and $\tau_a$. Due to the multiple resonant scattering of \lya\
photons on hydrogen and the concomitant increase of the photon path
length, $f_e$ is considerably smaller than the simple dust absorption
probability $\exp(-\tau_a)$, and the escape fraction decreases
strongly with increasing \nh.  

In an expanding shell the \lya\ photon destruction by dust depends
also to some extent on the gas temperature (or on the Doppler
parameter $b$), although in a somewhat ``subtle'' way as can be seen
from Fig.\ \ref{esc_frac_exp_shell}.  Indeed, for large column
densities (here \nh\ $> 10^{20}$) $f_e$ is found to depend little on
$b$, for intermediate values of \nh\ $\sim 10^{20}$ the escape fraction
varies in a non-monotonous way with $b$, and for lower \nh\ values
$f_e$ decreases with increasing $b$.  This latter behaviour is
opposite to the one in a static medium, where the escape fraction
increases with $b$ (cf.\ Fig.\ \ref{esc_frac_slab} or Eq.\ \ref{eq_fe}
with $a \propto b^{-1}$ and $\tau_0 \propto b^{-1}$).  The reason for
this inverted dependence is basically due to the fact that in an
expanding shell the quantity $\vexp/b$ determines the frequency at
which the initially emitted photons are seen by the atoms in the
receding shell. For increasing $b$, $\vexp/b$ decreases, so that
photons reaching the shell are seen with a frequency closer to line
center. Hence they will diffuse more and will have a higher
probability to be absorbed by dust. On the contrary, when $b$
decreases, photons are seen in the wings, their chance to interact
with dust decreases and so does $f_e$.  When they are sufficiently far
from line center ($\vexp/b \gg x_c$, where $x_c$ is some critical
frequency), the shell becomes transparent in \lya\ and the
escape fraction approaches the minimum value given by the dust 
absorption probability $\exp^{-\tau_a}$ (red upper
curve on Fig.~\ref{esc_frac_exp_shell}).

\subsection{Dust-free and dusty shells with arbitrary source spectra}

\begin{figure}[tb]
\includegraphics[height=6cm, width=8cm]{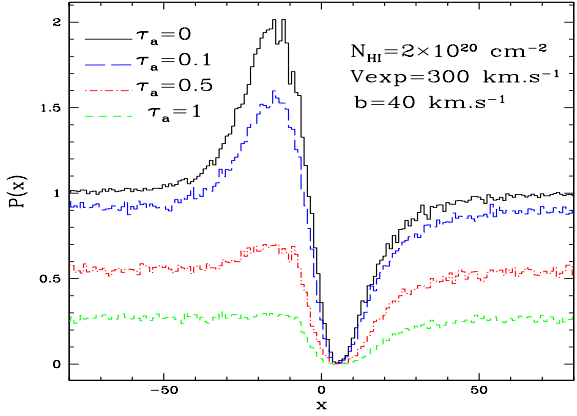}
\caption{Predicted \lya\ profiles for an expanding shell for a central source 
  emitting a flat continuum and with different amounts of dust. The
  adopted conditions are the following: 
$N_H=2. \, 10^{20}$ cm$^{-2}$, $T=10000$ K, $b=40$ \kms, and $\vexp=300$ \kms.
The line profiles are shown for dust amounts corresponding to
($M_d/M_H$, $\tau_a$) = (0., 0.), ($8.\, 10^{-4}$, 0.1),
($4.\, 10^{-3}$, 0.5), ($8.\, 10^{-3}$, 1.).}
\label{spec_cont_shell}
\end{figure}

Now we present the emergent spectrum from an expanding shell when the
input spectrum is a flat continuum and with different amounts of dust.
In Fig.\ \ref{spec_cont_shell}, the black solid line represents the
emergent spectrum from a dust-free shell. It is a
P-Cygni profile, quite similar to the expanding halo with a
central point source in Fig.\ \ref{dust_in_halos}. Again, \lya\
radiation transfer leads to the appearance of an emission peak and an
absorption feature in the emergent profile
which did not exist in the input spectrum.  The first remark is
that this spectrum is less complex than in the monochromatic case:
there is only one red peak, and no blue bump, due to radiation
transfer of photons at all frequencies. The emission peak is located at $x
\sim -2 x(\vexp)$, as the highest peak of the monochromatic spectrum,
and the absorption is around $x \sim x(\vexp)$, as this frequency
corresponds to the line center frequency in the shell frame. 
The effect of dust is similar to Fig.\ \ref{dust_in_halos} for
spherical halos: it suppresses the emission peak and broadens the
absorption, leading to an asymmetric absorption profile. The fraction
of escaping photons far from the line center is equal to
$\exp(-\tau_a)$, as expected (see discussion section~\ref{s_dust_in_halos}). 

\begin{figure}[tb]
\includegraphics[height=6cm, width=8cm]{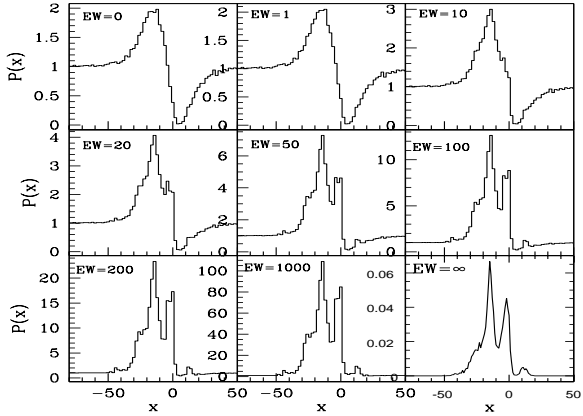}
\caption{Predicted \lya\ profiles from an expanding shell with a central source
emitting a flat continuum plus a monochromatic line (at \lya\ line center)
with different equivalent widths.
The physical conditions in the shell are: a \hi\ column density
  $N_{\rm \hi}=2\times 10^{20}$ cm$^{-2}$, a temperature $T=10^4$ K, a
  Doppler parameter $b=40$ \kms, no dust, and a radial velocity of
  $\vexp=300$ \kms. The \lya\ line equivalent width,
given in units of the Doppler width $\Delta\nu_D$, increases
  from $0$ to $\infty$ from the top-left to the bottom-right panel.}
\label{movie_shell}
\end{figure}

To illustrate the variety of emergent line profiles from a dust-free
expanding shell when varying the intrinsic line strength, we show in
Fig.\ ~\ref{movie_shell} a family of emergent spectra with intermediate
cases between intrinsic pure continuum
(cf.\ Fig.\ ~\ref{spec_cont_shell}) and intrinsic monochromatic
emission (cf.\ Fig.\  ~\ref{spec_exp_shell}). For a sufficiently
large \lya\ equivalent width  $W(\lya)$, the second red peak (feature
1a), the small excess at $x=0$ (feature 3) and the blue
bump (referred to as feature 2) become clearly visible.

\section{\lya\ line profile morphologies -- models and qualitative comparison
with observations}
\label{s_discuss}

For a better overview over the different simulations presented here and the resulting
variety of \lya\ line profiles, and for a first qualitative comparison with
observations we present a summary in Table \ref{t_overview}.

\begin{table*}
\caption{Schematic overview over the different simulations, the resulting line profile morphologies,
and observed objects showing qualitatively similar profiles.
Column 1 describes the adopted geometry, col.\ 2 describes the source position and spectrum,
col.\ 3 the predicted line profile morphology, col.\ 4 the corresponding figure in the 
paper, and col.\ 5 possible objects showing qualitatively similar profiles. }
\begin{tabular}{llllllll}
\hline
Case & Geometry  & Source &  Line profile & Figure  & Observations \\
\hline
\hline
1 & static medium & embedded source, & 2 symmetrical peaks & Fig.\ \ref{slab_spec} & LAB2?$^1$ \\
& & monochromatic & at $x_p=\pm(a\tau_0)^{1/3}$ & & \\	
2 & static medium & embedded source, & 2 symmetrical peaks if no dust & Fig. \ref{slab_cont} & \\
& & continuum & DLA profile not related to $\tau_0$ if dust &   & IZw18,SBS0335-052?  $^2$\\
3 & static medium & external source, & DLA profile & Fig.\ \ref{slab_cont_ext} & DLA$^3$\\
& & continuum & + faint diffuse component & \\
4 & expanding/infalling halo & central source, & asymmetric emission peak  & Figs.\ \ref{DLA}, \ref{movie_halo} \\
& & monochromatic &  & \\
5 & expanding/infalling halo & central source, & P-Cygni without dust & Figs.\ \ref{dust_in_halos}, \ref{movie_halo} \\
& & continuum & asymmetric absorption profile if dust   \\
6 & expanding shell & central source, & 1-2 red peaks (main peak at $\sim 2 \vmax$),  & Figs.\ \ref{spec_exp_shell}--\ref{exp_shell_N},
		\ref{movie_shell} & LBGs$^4$, low-$z$ starbursts$^2$\\
& & monochromatic & one blue peak  & \\
7 & expanding shell & central source, & P-Cygni without dust& Figs.\ \ref{spec_cont_shell}, \ref{movie_shell} & \\
& & continuum & asymmetric absorption profile if dust &  & IZw18,SBS0335-052?  $^2$\\ 
\hline
\multicolumn{7}{l}{{\em References:} $^1$ Wilman \etal\ (2005), $^2$ Mas-Hesse \etal\ (2003), $^3$ Adelberger \etal\ (2005),
	$^4$ see e.g.\ Shapley \etal\ (2003) and Noll \etal\ (2004)}   \\
\end{tabular}
\label{t_overview}
\end{table*}

Schematically we may classify the emergent \lya\ profiles and considered geometries in the 
three following main groups,
{\em 1)} static media and symmetric profiles,
{\em 2)} expanding/infalling halos, and
{\em 3)} expanding shells,
which we shall discuss now.

\subsection{Static media, symmetric profiles}
For simple static geometries with an embedded source emitting a
symmetric spectrum around $x=0$  (i.e.\ monochromatic line radiation,
a symmetric line centered on the \lya\ in the restframe of  the
medium, or a flat continuum) the emergent \lya\ line profile remains
symmetric with two  peaks, whose position is given by $x_p \approx \pm
(0.9-1) \times (a\tau_0)^{1/3}$  and little or no flux at line center
(case 1 in Table \ref{t_overview}; cf.\ Neufeld 1990, Hansen \& Oh 2006).   
The detailed line shape (FWHM of each part, the possible presence of
``bumps'' etc.) depends on the properties of the scattering medium in
ways quantified in detail by Neufeld (1990)  and examined also by
Richling (2003) for clumpy media. 
Breaking the symmetry of the source spectrum or introducing velocity
shifts between the source and the scattering medium lead to asymmetric
profiles, as well known (cf.\ Neufeld 1990). 

Emission peaks superposed to a continuum -- e.g.\ from a source with a
pure continuum  or continuum plus line emission -- are easily ``destroyed'' in
presence of dust (case 2). 
In this case the resulting profile is close to a Voigt profile.
However, due to radiation transfer effects and the assumed
geometry and ``aperture'', an integrated line profile fit  in this
manner will underestimate the true value of the column density. 

Purely or close to symmetric \lya\ line profiles are quite rarely observed. 
E.g.\ the spectra of the \lya\ blob LAB2 at $z \sim 3.09$ of Steidel
\etal\ (2000) observed by Wilman  \etal\ (2005), \lya\ emitters around
a radio galaxy (Venemans \etal\ 2005), or two LBG in the FORS Deep
Field (Tapken 2005) come close to this, and may hence be related to
static (or close to) media. For LAB2, however, other explanations have
been put forward by Wilman \etal\ (2005)  and Dijkstra \etal\ (2005b)
including outflows + ``absorbing screens'' and intergalactic medium
(IGM) inflow.  

Some nearby starbursts such as I Zw 18 and SBS 0335-052 show bf
``Voigt-like'' profiles (cf.\ Kunth \etal\ 1994, Thuan \& Izotov
  1997, Mas-Hesse \etal\ 2003). 
If related to a continuous source embedded in an \hi\ cloud (case 2), such profiles 
can be reproduced with sufficient dust. In this case the \hi\ column density
derived from simple Voigt profile fitting would underestimate the true value.
Other configurations can, however, also lead to the same profiles. 
To clarify such ambiguities, a detailed analysis taking all known constraints into 
account will be necessary.

A static medium illuminated by a distant background continuum source
(case 3) produces well known Voigt absorption line profiles, e.g.\
seen as damped \lya\ absorbers (DLA). 
In such a geometry \lya\ radiation transfer effects need not be considered.
A case of ``\lya\ fluorescence'' of QSO radiation on a nearby absorbing system
has been observed recently by Adelberger
et al.\ (2005): they detected a double-peaked emission superimposed on
a DLA profile. This can be related to our diffuse spectrum illustrated
in Fig~\ref{slab_cont_ext}.

\subsection{Expanding/infalling halos}
As already discussed by Zheng \& Miralda-Escud\'e (2002) and  Dijkstra
\etal\ (2005a) for monochromatic sources emitting at line center, such
geometries give rise to asymmetric profiles with a redshifted
(blueshifted) main peak for expanding (infalling) matter, and
eventually secondary features (case 4). The position of the main peak
shows the same dependence as for a Hubble outflow (Loeb \& Rybicky
1999): it moves away from line center with increasing $\nh$ and
decreasing velocity $ \vmax$. 
In the case of uniform
emissivity, the positions of the secondary peak and the point of
minimal flux indicate the external velocity (see Fig.\ 8 in Dijkstra \etal\ 2005a).

For sources emitting a pure continuum, the \lya\ radiation transfer leads to normal (inverted) 
P-Cygni profiles for expanding (infalling) halos. Again, as for static media, the presence
of dust is able to destroy the emission peak leading then to absorption line profiles
with more or less pronounced asymmetries (case 5).

Dijkstra \etal\ (2005b) have argued for possible infalling \lya\ halos in the case
of the \lya\ blob LAB2 of Steidel \etal\ (2000), and of a $z=5.78$ galaxy discovered 
by Bunker \etal\ (2003).
Numerous LBGs and LAEs show asymmetric redshifted \lya\ profiles, which
could in principle be related to expanding halos. To distinguish this geometry
from expanding shells or other geometries additional observational constraints 
are needed. For $z \sim 3$ LBGs, for example, expanding shells provide a good 
description, as we'll discuss now.
\subsection{Expanding shells}
Emergent \lya\ profiles from an expanding shell can be rather complex:
from P-Cygni profiles or asymmetric absorption profiles for an intrinsic
continuum spectrum (case 7) to emission line profiles made of one (or two)
red peaks and possibly a fainter blue bump for an intrinsically monochromatic 
(line) spectrum (case 6).
In all cases the profiles show an asymmetric red wing related to 
\lya\ scattering in the outflowing medium.
When \lya\ is in emission, the velocity shift of the main peak of the red part
of the profile corresponds to velocities between $\vexp$ and $\sim 2 \vexp$,
where  \vexp\ is the expansion velocity of the shell. For column densities
\nh\ $\ga 10^{20}$ cm$^{-2}$ the main red emission peak measures quite 
accurately twice the shell velocity.

Interestingly, spectroscopic observations of Lyman break galaxies
(LBG) at $z \sim 3$ agree very well with the expectations for an
expanding shell geometry.  Indeed the composite spectra of the large
LBG samples of Shapley \etal\ (2003) show not only clear signatures of
outflows, as testified by a shift between \lya\ emission and
interstellar absorption lines (with a mean value of $\Delta\nu_{\rm em
  - abs} \sim$ 650 \kms). The low-ionisation interstellar absorption
lines are found blueshifted by $\Delta\nu= - 150 \pm 60$ \kms\ with
respect to the stellar, systemic redshift, whereas the \lya\ emission
is found redshifted by $\Delta\nu= +360$ \kms\ (Shapley \etal\ 2003),
i.e.\ quite precisely at twice the velocity measured by the
interstellar (IS) absorption lines.  As the latter are formed by
simple absorption processes, i.e.\ by intervening gas on the line of
sight between the continuous source (the starburst) and the observer,
whereas \lya\ is prone to complex radiation transfer effects, the
above empirical finding together with our modeling insight indicates
very clearly that on average the distribution and kinematics of the
ISM in these LBGs is well described by an expanding shell.  This is
quite naturally the most simple geometry
to simultaneously explain 
IS absorption lines blueshifted by an outflow velocity \vexp\ and an
asymmetric \lya\ line redshifted by twice this velocity!

As these two spectral features are formed on opposite hemispheres, IS
absorption lines on the side approaching the observer and \lya\ from
photons scattered back from the receding shell (cf.\ Pettini \etal\
2001), this implies that
overall (or ``on average'' over an entire LBG sample) the ISM shell
structure must be fairly spherically symmetric rather than e.g.\
strongly bipolar.  Such a structure may indeed be expected in case of
strong starbursts triggering efficient large scale outflows (see e.g.\
simulations of Mori \etal\ 2002). 

For a shell geometry the finding of a velocity shift of $\sim 3 \vexp$ between
\lya\ and the IS absorption lines also requires a large enough \hi\
column density (cf.\ above). This is fully compatible with the
typical/average value of \nh\ of the $z \sim 3$ LBG sample of (Shapley
\etal\ 2003) expected from the standard correlation between \nh\ and
the extinction $E_{B-V}$. 

Finally we note that the relatively smooth, singly peaked \lya\ line
profiles observed by Shapley \etal\ (2003) and others are not in contradiction
with our predictions.  As already explained above, there are several
ways to avoid the multiple peaks predicted for certain conditions on
the red side of \lya\ (e.g.\ Figs.\ \ref{exp_shell_V} to \ref{exp_shell_N}):
among them are low expansion velocities, high column densities, or
large Doppler parameters and combinations thereof.  Actually, for the
relatively low outflow velocity of $\vexp \sim 150 \pm 60$ \kms\
deduced from the Shapley \etal\ (2003) sample, and taking into account
their relatively low spectral resolution, no distinguishable multiple
red peaks are expected. At least for this sample, the problem posed by
Ahn (2004) to avoid multiply peaked \lya\ profiles is ill posed.
The possible but generally minor blue peak, which is predicted but not
observed, does not pose a problem; in these high redshift objects it may
be too faint to be detected or actually suppressed by the intervening
interstellar \hi\ clouds in the galaxy. 

Local starbursts (see e.g.\ Kunth \etal\ 1994, 1998, Gonz\'alez Delgado 
\etal\ 1998, Mass-Hesse \etal\ 2003) present P-Cygni profiles with a faint 
secondary peak on the blue side superimposed to the absorption feature 
(e.g.\ IRAS 0833+6517), or more or less symmetric absorption profiles 
(e.g.\ IZw 18, SBS 0335-052). 
Mass-Hesse \etal\ (2003) interpret the variety of observed \lya\ profiles as 
phases in the time evolution of an expanding supershell. 
Few galaxies of those showing P-Cygni profiles have measurements 
of the relative velocities between the interstellar absorption lines, \lya,
and the ionised gas (traced e.g.\ by H$\alpha$). 
Although, e.g.\ in Haro 2 the shift between \lya\ and the interstellar lines
is $\sim 2 \vexp$, their spectra are a priori not in
contradiction with our simulations, as velocity shifts $< 3 \vexp$ are
obtained for neutral column densities in the shell smaller than
$10^{20}$ cm$^{-2}$, in agreement with the value derived by Lequeux
\etal\ (1995).
Furthermore, the
secondary peak on the blue side of the P-cygni profiles is easily
reproduced by \lya\ radiation transfer through a superbubble. This is
maybe the strongest indication for such a geometry.
Detailed modeling of these nearby galaxies, taking all available observational
constraints and spatial information into account, is necessary to shed more 
light onto these questions.

\subsection{General comments}
As the above shows, the expected \lya\ profiles show already quite a
complexity depending on the values of the main parameters considered
(geometry, \hi\ distribution and column density, velocity field,
Doppler parameter, dust to gas ratio, and input spectrum) and
important degeneracies are found. 
Overall it is quite clear from this study and earlier ones that
radiation transfer effects are important to shape the emergent \lya\ profile.
The interpretation of observed P-Cygni profiles or asymmetric emission
lines, for example, requires full radiation transfer calculations
rather than superpositions of gaussian emission and Voigt-like
absorption profiles as undertaken in various studies (e.g.\  Mas-Hesse
et al.\ 2003, Wilman et al.\ 2005).  Such a detailed comparison with
observations will be presented elsewhere. 

Obviously the geometries adopted in this paper are strongly idealised
when compared to real galaxies. 
Asymmetries,  non-unity filling or covering factors,
arbitrary geometries, and effects of different viewing and opening
angles have to be considered for more realistic simulations. 
Although our code can treat arbitrary density distributions, we have so far
only investigated homogeneous media. For complementary approaches treating
clumpy media see e.g.\ Neufeld (1991), and Hansen \& Oh (2006).
Furthermore, in addition to the simple limiting cases of input spectra
(monochromatic line plus continuum), other cases such as intrinsic Gaussian
emission lines must be accounted for.
All of these effects can easily be modeled with our generalised 3D radiation
transfer code, and will be discussed in future applications.

Beyond the \lya\ radiation transfer modeled here for simplified
``galaxian'' geometries it should be recalled that \lya\ line profiles
predicted in this manner are still a priori prone to alterations due
to the transmission through the intervening IGM., 
As well known (e.g.\ Madau 1995) this effect can significantly alter 
observed \lya\ line profiles --- especially on the blue side --- for objects 
of sufficiently high redshift (typically $z \ga\ 3$). Effects of the IGM
have e.g.\ been discussed by Haiman (2004), Santos (2004), and
Dijkstra \etal\ (2005b), and are not treated here.
For most purposes the IGM effects can be computed without the need for true
\lya\ radiation transfer calculations.

\section{Summary and conclusion}
\label{s_conclude}
We have developed a new general 3D \lya\ radiation transfer code
allowing for arbitrary hydrogen density, ionisation \& temperature
structures, dust distributions, and velocity fields using a Monte
Carlo technique (Sect.\ \ref{s_code}). 
Currently the code works on a cartesian grid and it can handle
\lya\ transfer problems in objects with arbitrary hydrogen column densities.
The main radiation--matter interaction processes
included in the code are \lya\ scattering, dust scattering, and dust
absorption.  UV/\lya\ photon sources with an arbitrary spatial
distribution and arbitrary spectra can be treated.
The main direct observables predicted by our simulations are ``global'' 
(i.e.\ integrated over all viewing angles) \lya\ line profiles, 
spectra for different viewing directions and opening angles, and
surface brightness maps at different wavelengths (or integrated over \lya).

The code has been tested successfully and compared against analytical results
and results from other simulations (e.g.\ Neufeld 1990, Ahn \etal\ 2001, 
Zheng \& Miralda-Escud\'e 2002, Richling \etal\ 2003, Dijkstra \&
Haiman 2005a) for various geometrical configurations
including plane parallel slabs, disks, and expanding or infalling halos,
and for computations with and without dust.
Overall an excellent agreement is found, except in cases where assumptions
made in the analytical calculation break down (see Sect.\ \ref{s_test}).

With the aim of understanding and interpreting ultimately the variety of 
\lya\ profiles observed in nearby and distant starbursts, we have examined 
the \lya\ line profiles predicted for several simple geometrical
configurations and their dependence on the main parameters such as the
\hi\ column density, temperature, velocity, and dust content.  
We have considered slabs with internal sources, disks, expanding and
infalling halos (Sect.\ \ref{s_test}), externally illuminated slabs
(Sect.\ \ref{s_slab}), and expanding shells (Sect.\ \ref{s_shell}).  
For the source spectrum we have considered the limiting cases of pure
monochromatic line emission, a pure continuum, or intermediate cases
with different line strengths. The two latter cases have rarely
been discussed in the literature so far.

Schematically, the following morphologies are found
(cf.\ Table \ref{t_overview}): 
profiles reaching from doubly peaked symmetric emission to symmetric
 "Voigt-like" (absorption)
profiles in static configurations with increasing dust content, and
asymmetric red- (blue-) shifted emission lines with a blue (red) counterpart
ranging from absorption to emission (with increasing line/continuum strength)
in expanding (infalling) media.
In principle, symmetric or nearly symmetric profiles are only obtained
for static geometries or for small systematic velocities. In practice,
however, this may be  altered by \lya\ scattering/absorption in the
intervening IGM, especially at high  redshift.

Some specific results are worthwhile summarising here:
\begin{itemize}
\item Pure \lya\ absorption line profiles may be observed in
  ``integrated''/global spectra  of objects with static geometries. In
  this case the naturally arising double \lya\ emission peaks have
  been suppressed by dust absorption. Standard Voigt profile fitting
  of such profiles will significantly underestimate the true hydrogen
  column density (see Sect.\ \ref{s_slab}).

\item It should be noted that normal (inverted) P-Cygni like \lya\ profiles
can be obtained in expanding (infalling) media from objects without
any intrinsic \lya\ emission, as a natural consequence of radiation
transfer effects redistributing UV (continuum) photons from the \lya\
line center to the red (blue) wing (e.g.\ Figs.\ \ref{movie_halo},
\ref{movie_shell}). 
Adding dust in such cases will progressively transform the P-Cygni profile
into a broad asymmetric absorption profile.

\item Our 3D simulations of expanding shells lead us to revise the
  earlier results of Ahn and collaborators (Ahn \etal\ 2003, Ahn 2004)
  and to clarify the radiation transfer processes explaining the
  detailed shape of \lya\ profiles predicted from expanding shells. In
  particular we found that, for sufficiently large column densities
  (\nh\ $\ga 10^{20}$ cm$^{-2}$), the position of the main \lya\
  emission peak is quite generally redshifted by approximately twice
  the expansion velocity (Sect.\ \ref{s_shell}). 
This is in excellent agreement with the observations of $z \sim 3$ LBGs
from Shapley \etal\ (2003), which show that \lya\ is redshifted
by  $\sim 2 \vexp$, where $\vexp$ is the expansion velocity measured from the
interstellar absorption lines blueshifted with respect to the stellar (i.e.\
galaxian) redshift.
This finding indicates also that large scale, fairly symmetric shell
structures must be a good description for the outflows in LBGs.
\end{itemize}

As already clear from earlier investigations, radiation transfer
effects even in the simple configurations just mentioned give rise to
a complex morphology of \lya\ line profiles
and some degeneraciesare found e.g.\ between the main parameters such
as the hydrogen column density \nh, the Doppler parameter $b$, and
also the intrinsic emission line width. 
However, codes as the one developed and presented here allow one to undertake
detailed \lya\ line profile fitting to determine the physical
properties of the  emitting gas and of the surrounding  \hi\ and dust
and to quantify possible degeneracies. With such tools it is now
possible to fully exploit the information encoded in \lya\ profiles
without recurring to physically  unmotivated superpositions of
emission and Voigt absorption line profiles, which -- except in
specific geometric configurations -- generally do not properly account
for radiation transfer effects.
Such detailed studies of nearby and distant starforming galaxies will
be presented in subsequent publications. 
Various other applications, combining also our {\em MCLya} code with the 3D 
radiation transfer and photoionisation code CRASH (Maselli \etal\
2003) are foreseen.

\acknowledgements

AV thanks the Arcetri Observatory and its staff for their hospitality
during a stay.
We thank Andrea Ferrara, Miguel Mas-Hesse, Ruben Salvaterra, Christian
Tapken, and David Valls-Gabaud for useful discussions and comments on an
earlier version of the paper.
We thank Daniel Kunth and the other participants of the ``\lya\ team'', 
Miguel Mas-Hesse, G\"oran \"Ostlin, and Matthews Hayes,
for lively discussions and confrontations with observations
at a team meeting.
This work was supported by the Swiss National Science Foundation. 
AM is supported by the ``DFG Priority Program 1177''. 

\end{document}